\documentclass[
aip,
amsmath,amssymb,
preprint
]{revtex4-1}

\usepackage{graphicx}
\usepackage{dcolumn}
\usepackage{bm}

\usepackage[utf8]{inputenc}
\usepackage[T1]{fontenc}

\usepackage{footnoterange}
\usepackage{amsmath}
\usepackage{amsthm}
\usepackage{amssymb}
\usepackage{mathrsfs}
\usepackage{mathtools}
\usepackage{epsfig}
\usepackage{caption}
\usepackage{subcaption}

\usepackage{color}
\usepackage{wasysym}
\usepackage{booktabs}
\usepackage{multirow}
\usepackage{siunitx}
\usepackage{natbib}

\usepackage{epstopdf}
\usepackage[dvipsnames]{xcolor}
\usepackage{tikz}
\usepackage{tabularx}
\usepackage{ragged2e}

\usepackage{pifont}
\usepackage{xcolor}
\usepackage{hhline}
\usepackage{scalerel,stackengine}

\newcommand\reallywidehat[1]{\ensurestackMath{%
\savestack{\tmpbox}{\stretchto{%
\scaleto{%
\scalerel*[\widthof{\ensuremath{#1}}]{\kern-.6pt\bigwedge\kern-.6pt}%
{\rule[-\textheight/2]{1ex}{\textheight}}
}{\textheight}%
}{0.5ex}}%
\stackon[1pt]{#1}{\tmpbox}%
}}

\usepackage[colorlinks,
            colorlinks=true,
            linkcolor=blue,
            citecolor=blue,
            filecolor=magenta,
            urlcolor=black]{hyperref}
    
\graphicspath{{figures/}}

\begin{document}

\preprint{AIP/123-QED}

\title{A Data-Driven Dynamic Nonlocal Subgrid‐Scale Model for Turbulent Flows}

\author{S. Hadi Seyedi}
\affiliation{Department of Mechanical Engineering, Michigan State University, 428 S Shaw Ln, East Lansing, MI 48824, USA}
\author{Mohsen Zayernouri} \email{zayern@msu.edu}
\affiliation{Department of Mechanical Engineering, Michigan State University, 428 S Shaw Ln, East Lansing, MI 48824, USA}
\affiliation{Department of Statistics and Probability, Michigan State University, 619 Red Cedar Road, Wells Hall, East Lansing, MI 48824, USA}

\date{\today}

\begin{abstract}
We developed a novel autonomously dynamic nonlocal turbulence model for the large and very large eddy simulation (LES, VLES) of the homogeneous isotropic turbulent flows (HIT). The model is based on a generalized (integer-to-noninteger) order Laplacian of the filtered velocity field, and a novel dynamic model has been formulated to avoid the need for tuning the model constant. Three data-driven approaches were introduced for the determination of the fractional-order to have a model which is totally free of any tuning parameter. Our analysis includes both the \textit{a priori} and the \textit{a posteriori} tests. In the former test, using a high-fidelity and well-resolved dataset from direct numerical simulations (DNS), we computed the correlation coefficients for the stress components of the subgrid-scale (SGS) stress tensor and the one we get directly from the DNS results. Moreover, we compared the probability density function of the ensemble-averaged SGS forces for different filter sizes. In the latter, we employed our new model along with other conventional models including static and dynamic Smagorinsky into our pseudo-spectral solver and tested the final predicted quantities. The results of the newly developed model exhibit an expressive agreement with the ground-truth DNS results in all components of the SGS stress and forces. Also, the model exhibits promising results in the VLES region as well as the LES region, which could be remarkably important for the cost-efficient nonlocal turbulence modelings e.g., in meteorological and environmental applications.
\end{abstract}

\maketitle

\section{Introduction}\label{sec: Intro}
The prohibitively high computational cost of direct numerical simulations (DNS) of realistic turbulent flows has motivated the community of research in turbulence to develop coarse-grained techniques including Reynolds-averaged Navier–Stokes (RANS) and large-eddy simulation (LES) methods to reduce the intractably large degrees of freedom in DNS studies\cite{pope2001turbulent, sagaut2006large, meneveau2000scale, germano1991dynamic, moin1991dynamic, vasilyev1998general}. Using Reynolds averaging approach in the Navier–Stokes (N-S) equations provide temporally averaged quantities, however, in LES approaches one employs a subgrid-scale (SGS) model, which represents the effects of the finer scales\cite{pope2004ten, pope2001turbulent}.

Turbulence experimental and DNS features have confirmed that the turbulence is intrinsically nonlocal, and its statistics are non-Gaussian, which means velocity increments have sharp peaks, heavy-skirts and also skewed\cite{wilczek2017emergence, akhavan2020anomalous}. However, most of the turbulence models have been built based on the Boussinesq's turbulent viscosity concept, in which one assumes turbulent stress tensor is proportional to the \textit{local} mean velocity gradient at any point, and the proportionality coefficient is set to the turbulent viscosity. Prandtl in 1942 aimed to disregard this local constraint by introducing the \textit{extended mixing length} concept for the first time. The new model was a migration from locality to nonlocality, however, the model and its implementation were not remarkably successful since the scale of nonlocality was comparable with the differential length scale. Afterward, Prandtl parametrized the model in a way that the mixing length was taken to be bigger than the differential length. This strategy was the same as adding a weak nonlocal concept to the model, hence called \textit{weak nonlocal}. Bradshaw \cite{bradshaw1973agardograph}  in 1973 showed that Boussinesq's hypothesis fails over curved surfaces and noted that form of the stress-strain relations is responsible for this failure. It should be mentioned that there were some important developments mostly based on polynomial series compared to the Boussinesq type modeling including the works done by Spencer and Rivlin \cite{spencer1958theory, spencer1959further}, Lumley \cite{lumley1970toward} and Pope\cite{pope1975more}; however, they all lacked the accuracy that a ``true'' physical modeling should provide especially for the second-order and higher tensor series development.

Using generalized-order derivatives is a relatively convenient approach to bring in nonlocality concept from mathematical point of view. The generalized-order operators represent the underlying heavy-tailed stochastic processes at the continuum level, which can be properly utilized in incorporating the long-range interactions in various mathematical models including but are not limited to beam vibration analysis \cite{suzuki2021anomalous}, anomalous rheology modeling \cite{suzuki2021data, naghib2018ear,naghibolhosseini2015estimation},
damage modeling considering memory effects \cite{suzuki2021thermodynamically} and visco-elasto-plastic models \cite{suzuki2016fractional}. Moreover, harnessing the generalized-order models capabilities can be obtained properly using the highly accurate numerical schemes for integer and fractional-order PDEs \cite{d2020numerical, kharazmi2019fractional, samiee2019unified, zhou2020implicit, seyedi2019high, lischke2019spectral, seyedi2018multiresolution, zayernouri2014fractional, kharazmi2017petrov, zayernouri2013fractional, kharazmi2018fractional}, which is also an active research topic.

In a pioneer work by  Hinze et al. \cite{hinze1974memory} in 1974, the authors described the memory effect in a turbulent boundary layer flow. They used the experimental data, produced downstream of a hemispherical cap attached to the lower wall of channel geometry and illustrated that, when one computes eddy-viscosity using Boussinesq's theory in the lateral gradient of the mean flow, there is a significant non-uniform distribution that also exists in the outer region of the boundary layer. Interestingly, a nonlocal expression for the gradient of the transported field was proposed in a novel approach by Kraichnan in the same year for the scalar quantity transport  \cite{kraichnan1964direct}. Afterward, fractional-order models based on the RANS approach were offered in \cite{chen2006speculative, epps2018turbulence, egolf2017fractional, hamba1995analysis, hamba2005nonlocal}. One of the main contributions in the development of nonlocal RANS models is by Egolf and Hutter \cite{egolf2017fractional, egolf2020nonlinear}. They started from L\'evy flight statistics and generalized the zero-equation local Reynolds shear stress expression to a nonlocal and fractional type. The method is based on Kraichnanian convolution-integral approach and utilizing different weighting functions. Using the mentioned weighting functions, one can make a bridge between the first-order gradient of the common eddy diffusivity models and the mean velocity difference term. Their proposed model is called the Difference-Quotient Turbulence Model (DQTM) and is based on the four distinct steps that can be followed to change a local operator to a nonlocal one. In reality, the proposed model is a more general version of Prandtl's zero-equation mixing length and shear-layer turbulence models.

There is also emerging attention for the nonlocal LES models. Samiee et al. proposed a new model for the HIT flows based on the fractional Laplacian by employing L\'evy stable and tempered L\'evy stable distributions in kinetic level \cite{samiee2020fractional, samiee2021tempered}. They showed that the new nonlocal models can recover the non-Gaussian statistics of subgrid-scale stress motions. Laval et al. \cite{laval2001nonlocality}  analyzed the effects of the local and nonlocal interactions on the intermittency corrections in the scaling properties. They observed that nonlocal interactions are responsible for the creation of the intense vortices and on the other hand, local interactions are trying to dissipate them. Akhavan-Safaei et al. \cite{akhavan2021data}  proposed a fractional LES approach for the subgrid-scale modelings of the scalar turbulence. They utilized the two-point statistics for defining the optimal fractional-order of the new nonlocal model, and by using \textit{a priori} assessment they showed that there is proper agreement between the probability distribution function (PDF) of the SGS dissipation and the one that comes from the filtered DNS data. Harmonious with this study, Akhavan-Safaei and Zayernouri developed a corresponding nonlocal spectral transfer model and a new scaling law for scalar turbulence in \cite{akhavan2021nonlocal-transfer}. Their new analysis additionally reconciled the close similarities between this work and their earlier development in \cite{akhavan2021data} when the filter scale approaches the dissipative scales of turbulent transport. There are also other related studies that one can consult with, including preliminary fractional modeling in wall-bounded turbulent flows \cite{keith2021fractional}, \textit{a priori} survey of nonlocal eddy viscosity-based model for the isotropic and anisotropic (channel flow canonical test cases) turbulent flows \cite{di2021two}, hybrid nonlocal model in the case of magnetically confined plasma \cite{milovanov2014mixed}, generalization of a deconvolution model with fractional regularization for the rotational N-S equations \cite{ali2014theory}, and fractional Laplacian closure and its connection to Richardson pair dispersion \cite{gunzburger2018analysis}. Going even beyond the scope of research in turbulence, a new comprehensive survey on the nonlocal models for several crucial applications, including anomalous subsurface dynamics, turbulence modeling, and extraordinary materials, was recently performed by Suzuki et al. \cite{suzuki2021fractional}.

Considering the nonlocal models in the literature, there are some important imperfections including the sensitivity to the model constant and fractional-order parameter, relatively low correlation coefficients $\rho ( \tau ^{\Delta},  \tau ^{\Delta, Model} ) $, and no back-scatter prediction of kinetic energy from small scales to large scales. However, the conventional and frequently utilized local LES turbulent models are being improved over time to be free from mentioned deficits. One of the methods in the improvement process is using the \textit{dynamic} procedure for the determination of the model constant \cite{germano1991dynamic}.

To fill the gap in the literature and provide an applicable and relatively easy to implement nonlocal LES model, we have developed a new dynamic nonlocal model that accounts for all the aforementioned downsides. In the new dynamic fractional subgrid-scale model (D-FSGS), both nonlocality and dynamic features have been leveraged together for the first time. This match between two important features provides a unique and higher performance than the dynamic local or static nonlocal models. Interestingly, the analysis showed that in the new model, we have remarkably less sensitivity to the fractional-order, which is needed to be specified in nonlocal models. This relative freedom is obtained thanks to the novel coupling between the dynamic procedure and the nonlocal nature of the base model. In the following, we derived and implemented the D-FSGS model in both \textit{a priori} and \textit{a posteriori} stringent tests and compared the results with the conventional local models including Smagorisnky (SMG) and dynamic Smagorinsky (D-SMG) models.

This study is organized as follows: in section \ref{sec: Model_Development} we talk about the governing equations and development of the new model. Section \ref{sec: APriori_Analysis} starts with introducing three main distinct approaches for the determination of the fractional-order, and then we do a comprehensive \textit{a priori} assessment along with a comparative study with conventional models. In section \ref{sec: APosteriori_Analysis}, we study the numerical stability of the proposed model and do different tests including two-point diagnostics to have a complete overview of the model performances in a \textit{a posteriori} sense. Finally, in section \ref{sec: Conclusion} we recap the findings with a conclusion.
\section{Model Development}\label{sec: Model_Development}
Implementation of a low-pass filter on the N-S equation forms a closure term on the right-hand side of the momentum equation, which needs to be modeled, representing the (unknown) SGS dynamics. The filtered incompressible N-S equations can be written as
 \begin{align}\label{eqn: NS}
\frac{\partial \bar{u}}{\partial t} + \bar{u} \cdot \nabla \bar{u} &= - \frac{1}{\rho} \ \nabla \bar{p} + \nu  \  \nabla ^{2} \bar{u} - \nabla \cdot \tau, \\ \nonumber
    \nabla \cdot \bar{u} &= 0.
\end{align}
In this equation, $\bar{u}$ is the filtered velocity vector, $\rho$ denotes the constant density, $\bar{p}$ represents the filtered pressure, and $\nu$ shows the kinematic viscosity. The effects of the small scales arise in the so-called SGS stress tensor 
\begin{align}\label{eqn: SGS}
 \tau_{ij} = \overline {u_{i}u_{j}} - \bar{u_{i}}\bar{u_{j}},
\end{align}
forming the closure term to be modeled. Several models have been proposed during the last decades to close the filtered N-S equation in both functional and structural LES models \cite{pope2001turbulent, meneveau2000scale, sagaut2006large}. In the classical Smagorinsky model (SMG)\cite{smagorinsky1963general},  the deviatoric part of the stress tensor is written as
\begin{align}\label{eqn: SMG}
 \tau_{ij}^{SMG} = - 2 \ \nu_{t} \ \bar{S_{ij}},
\end{align}
where $\nu_{t}$ denotes the eddy viscosity, and $\bar{S_{ij}} = \frac{1}{2} ( \frac{\partial \bar{u_i}}{\partial x_j} +  \frac{\partial \bar{u_j}}{\partial x_i} )$ represents the filtered (resolved) strain rate tensor. In this model, $\nu_{t}$ is constructed based on the Prandtl's Mixing Length hypothesis,
\begin{align}\label{eqn: nut}
 \nu_{t} = C_{s} \ \mathcal{L}^{2} \ |\bar{S}|,
\end{align}
where $\mathcal{L}$ shows the effective grid scale, and $|\bar{S}| = \sqrt{2 \bar{S_{ij}} \bar{S_{ij}}}$ exhibits the magnitude of the resolved scale strain rate tensor \cite{meneveau2000scale}. One of the main drawback of this model is its flow-dependent feature, which means that the model can not correctly predict final quantities with a single universal constant in different scenarios such as shear flows, wall-bounded flows, or transitional flows. To overcome this challenge in the Smagorisnky model, Germano et al. \cite{germano1991dynamic} proposed a novel procedure for evaluation of the model coefficient, which is called the \textit{dynamic Smagorinsky} model (D-SMG). The suggested procedure was a breakthrough in turbulence modeling, and several researchers utilized the same concept afterward \cite{lund1995experiments, najjar1996study, piomelli1995large, ghosal1997numerical}. The dynamic procedure was designed based on the classic idea that one can extract useful information from the smallest resolved scales for modeling the subgird-scales; however, the way it was applied was indeed novel. It calculates the eddy-viscosity coefficient locally for each LES grid point as the calculation progresses, and there is no need for any predefined inputs in the model. This model was constructed based on the scale-invariance hypothesis and calculates the model constant using the information from the resolved section.

\subsubsection{Fractional SGS Model \ \cite{samiee2020fractional}: }\label{subsec: FSGS }
We have recently developed a nonlocal SGS model that we present here briefly for making this work self-contained. Starting from the Boltzmann kinetic level description of the flow, we employ the BGK model for the collision of the particles and have
\begin{align}\label{Bolt}
\frac{\partial f}{\partial t} + u \cdot \nabla f = - \frac{f-f^{eq}}{\mathcal{T}},
\end{align}
where $f = f(x,u,t) $ is called the single-particle probability distribution function and shows the particles density in the phase space $(x ,u)$ at time $t$. Moreover, $f^{eq}$ represents the local equilibrium distribution function and is defined based on the Maxwell distribution.
\begin{align}\label{Maxwell}
f^{eq}(\mathcal{K}) = \frac{\rho}{U^3}F(\mathcal{K}),
\end{align}
where $F(\mathcal{K})=e^{-\mathcal{K}/2}$, $\mathcal{K}=\frac{\vert {u}-V\vert^2}{U^2}$ and ${U}$ corresponds for the thermal agitation speed. The left-hand side of Eq. \ref{Bolt} correspond to the streaming of the non-reacting particles and right-hand side shows the collision operator with a relaxation time $\mathcal{T}$. One can solve Eq. \ref{Bolt} analytically by method of characteristic to find distribution in terms of equilibrium state \cite{chen2010macroscopic}
\begin{align}\label{Bolt_solution}
f (t,x, u) =\int_{0}^{\infty} e^{-s} \,  {f^{eq}(x-u\tau s, u, t-\tau s)}\, ds.
\end{align}
Moments of the $f$ would provide the macroscopic flow variables. Therefore, one can write $\rho = \int f(t,x,u) du$, and $\rho V(t, x) = \int u f(t,x,u) du$ to compute density and fluid velocity.
Incorporating filtering procedure into the Eq. \ref{Bolt} would provide filtered Boltzmann equation as
\begin{align}\label{F_Bolt}
\frac{\partial \bar{f}}{\partial t} + u\cdot \nabla \bar{f} = - \frac{\bar{f}-\overline{f^{eq}(\Delta)}}{\mathcal{T}}.
\end{align}
However, the collision term is highly-nonlinear and filtering kernel cannot commute \cite{girimaji2007boltzmann}. Hence, a closure problem would be built by defining $\overline{\mathcal{K}}=\frac{\vert {u}-\overline{V}\vert^2}{U^2}$ since
\begin{align}\label{inequality}
\overline{f^{eq}(\mathcal{K})}   \neq  f^{eq}(\overline{\mathcal{K}}).
\end{align}
One of the approaches for handling this problem is using a power-law distribution for modeling $\overline{f^{eq}(\mathcal{K})}  - f^{eq}(\overline{\mathcal{K}})$, i.e., 
\begin{align}\label{model-powerlaw}
\overline{f^{eq}(\mathcal{K})}-  f^{eq}(\overline{\mathcal{K}}) \simeq f^{Model}(\bar{\mathcal{K}}) =  \mathfrak{D}_{\beta}\, f^{\beta}(\bar{\mathcal{K}}),
\end{align}
in which $f^{\beta}(\bar{\mathcal{K}}) = \frac{\rho}{U^3}F^{\beta}(\mathcal{K})$, where $F^{\beta}(\mathcal{K})$ represents an isotropic \textit{L\'evy} $\beta$-stable distribution. Also, $ \mathfrak{D}_{\beta}$ is a constant which is going to be addressed in the present dynamic model. In the fractional Laplacian model, the SGS forces are defined as 
\begin{align}\label{eq:force}
           (\nabla . \tau)_{i}  = \nu_{\alpha} (-\Delta ^ {\alpha}) \bar u_{i}, \quad \alpha \in (0,1],
\end{align}
where the filtered velocity $\bar u_{i}:\mathbb{R}^d\times (0,T] \rightarrow\mathbb{R}$, where $d=3$ is the dimension of physical domain and $T$ represents the simulation time, in addition, $ (-\Delta ^ {\alpha}) (\cdot)$ denotes the space-fractional Laplacian of  order $2\alpha \in (1,2]$, which can be defined as a \textit{singular integral operator} given by
\begin{align}\label{eq:fractional-derivative}
(-\Delta)^{\alpha} \bar u_{i} = c_{d, \alpha}\int\limits_{\mathbb{R}^d}{\frac{\bar u_{i}(\bm{x})-\bar u_{i}(\bm{y})}{|\bm{x}-\bm{y}|^{d+2\alpha}}\,d\bm{y}},
\end{align}
where $c_{d, \alpha} = \frac{4^\alpha\Gamma(d/2+\alpha)}{\pi^{d/2}|\Gamma(-\alpha)|}$. Moreover employing the periodic boundary conditions, the corresponding Fourier transform of the fractional Laplacian in \eqref{eq:fractional-derivative} is given by (see e.g., \cite{lischke2020fractional})
\begin{align}\label{eq:Fourier}
\mathcal{F} \Big [ (-\Delta ^ {\alpha}) \bar u_{i}(\bm{x}) \Big] = |\bm{k}|^{2\alpha} \mathcal{F} [\bar u_i] (\bm{k}),
\end{align}
in which $\mathcal{F}$ and $\bm k$ are the Fourier transform and Fourier numbers, respectively. Evidently, the integer-order Laplacian operator is recovered simply by putting $\alpha = 1$. The Fourier transform \ref{eq:Fourier} provides a rather convenient way of handling the fractional operators Fourier space and in our Fourier spectral method for simulating the problem. The corresponding ``eddy-viscosity"-like model coefficient in \eqref{eq:force} is then obtained as
\begin{align}\label{eq:nu }
          \nu_{\alpha} = \mathcal{C} \,F(\alpha),
\end{align}
where $\mathcal{C}$ represents an up-scaling model input being proportional to $ U^{^{2\alpha}} \mathcal{T} ^{2\alpha -1} $ from the kinetics theory's perspective, yet to be dynamically computed in the subsequent (continuum-level) simulations, moreover, $F$ denotes a deterministic univariate function of the fractional-order $\alpha$, explicitly given by
\begin{align}\label{eq:F_alpha }
          F(\alpha) =          \frac{ 2^{2 \alpha} \ \Gamma (\frac{2 \alpha +3 }{2}) \,\, \Gamma (2 \alpha +1)  } {\pi ^{\frac{3}{2}}  |\Gamma (-\alpha)|},
\end{align}
rendering \eqref{eq:force} as $(\nabla . \tau)_{i}  = \mathcal{C} \, F(\alpha)\, (-\Delta ^ {\alpha}) \bar u_{i}, \, \alpha \in (0,1]$. In what follows and in this generalized order context, we develop a new dynamic procedure to automatically compute $\mathcal{C}$ from data on-the-fly.

\subsubsection{Derivation of the Nonlocal Dynamic Model Procedure:}\label{subsec: Dynamic Model Procedure}
We write $(\nabla . \tau)_{i}  = C_1 F(\alpha) (-\Delta ^ {\alpha}) \bar u_{i}$ for the sake of simplicity. Implementing the second filtering process (test-level filter), gives the divergence of the SGS stresses at the test-level filter (subtest-scale stress) as
\begin{align}\label{eq: }
          (\nabla . T)_{i}  = C _{2} F(\alpha) (-\Delta ^ {\alpha}) \widehat{\bar u_{i}} .
\end{align}
In the above equation, $\widehat{(\cdot )}$ indicates the test-level filtering, which is commonly chosen as twice the grid-level filtering. Now, the Germano identity \cite{germano1991dynamic}, which relates the stresses at grid-level $(\tau_{ij})$ and test-level $(T_{ij})$, is employed to make a bridge between the resolved scales and the subgrid-scales,
\begin{align}\label{eq: Germano }
          G_{ij} =\widehat{ \bar u_i \bar u_j} -  \widehat{ \bar u_i } \widehat{ \bar u_j } = T_{ij} - \widehat{\tau _{ij}}, 
\end{align}
which is a known quantity, and represents the resolved turbulent stress. In the divergence form, we have
\begin{align}\label{eq: }
        \nabla .\big(  \widehat{ \bar u_i \bar u_j} -  \widehat{ \bar u_i } \widehat{ \bar u_j } \big) = (\nabla . G )_{i}. 
\end{align}
Therefore, we construct the Germano identity in the divergence form based on the previously introduced fractional Laplacian model. The Germano identity extrapolates and parametrize the model constant for the subgrid-scale part using the information in the smallest resolved scales,
\begin{align}\label{eq.tensorial}
     (\nabla . T)_{i} - (\nabla . \widehat{ \tau})_{i}  = C \bigg(  F(\alpha ) (-\Delta ^ {\alpha}) \widehat{\bar V_{i}} - \reallywidehat{ F(\alpha ) (-\Delta ^ {\alpha}) \bar V_{i}} \bigg). 
\end{align}
Here, we assume the model constant $C$ being scale invariant around the grid and test filter size, ($C = C_1 = C_2$), and $C$ is in general a space-time dependent tensor, yet, it has been assumed to be spatially uniform. One needs to solve Eq. \ref{eq.tensorial} for $C$, which forms a tensorial equation. There are two common methodologies to get a scalar model constant. First, one may contract with $\bar{S_{ij}}$ as mentioned originally in \cite{germano1991dynamic}. Second approach is proposed by Lilly \cite{lilly1992proposed}, which is more preferable and commonly used. This approach is based on contracting with the tensor, which has been multiplied to the unknown coefficient in the right hand side of  Eq. \ref{eq.tensorial}.
 \begin{figure}[t!]
    \begin{minipage}[b]{.49\linewidth}
        \centering
        \includegraphics[width=\textwidth]{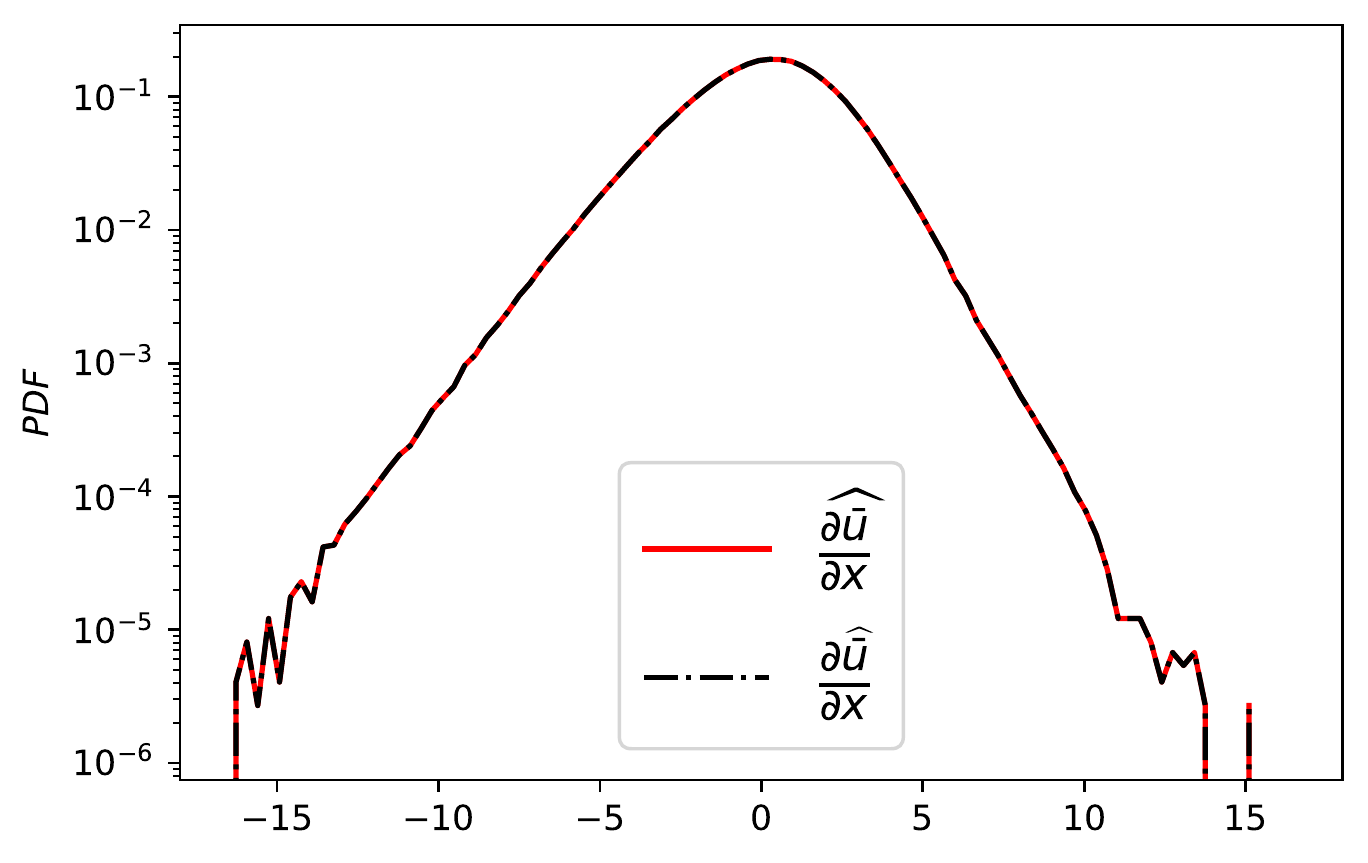}
        \subcaption{}\label{fig: Commute1}
    \end{minipage}
    \begin{minipage}[b]{.49\linewidth}
        \centering
        \includegraphics[width=\textwidth]{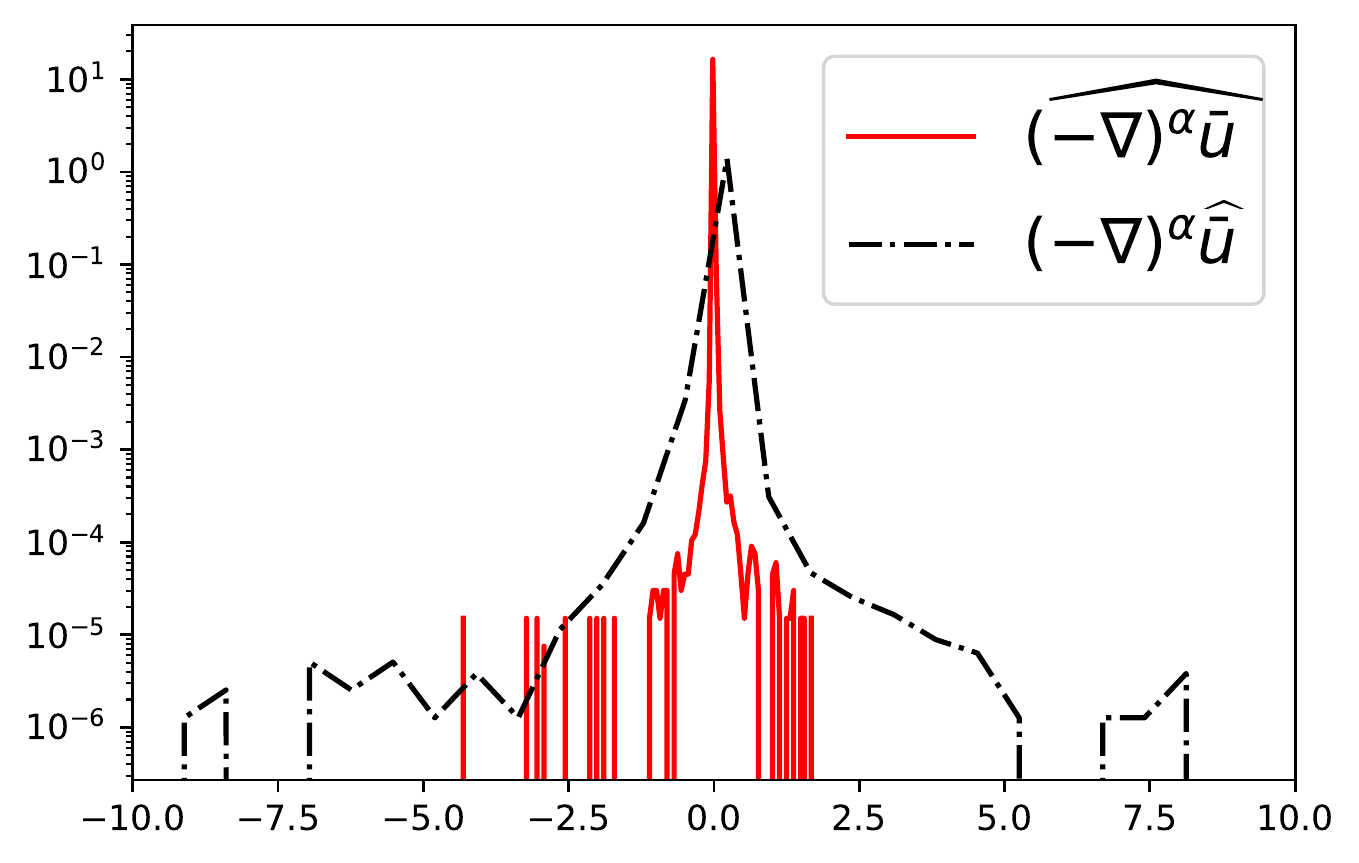}
        \subcaption{}\label{fig: Commute2}
    \end{minipage}
    \caption{Difference between commuting of the low pass filter on the (a) integer order derivatives, and (b) fractional-order derivative ($\alpha = 0.6$).}\label{fig: Commute}
\end{figure}
One important point that should be mentioned in this section is that in the conventional local LES modeling, the main assumption is that the operation of filtering and integer-order differentiation commutes, then we get a set of filtered equations. Commuting means commutation between the filtered quantity and the spacial derivatives. However, the filtering procedure does not commute with the fractional-order operators in general. One simple example regarding this important matter is depicted in Figure \ref{fig: Commute} for $\alpha = 0.6$, comparing the derivatives of the filtered velocity at test-level and filtered derivative for the integer and fractional-order cases.

\begin{figure}[t!]
        \centering
        \includegraphics[width=0.6\textwidth]{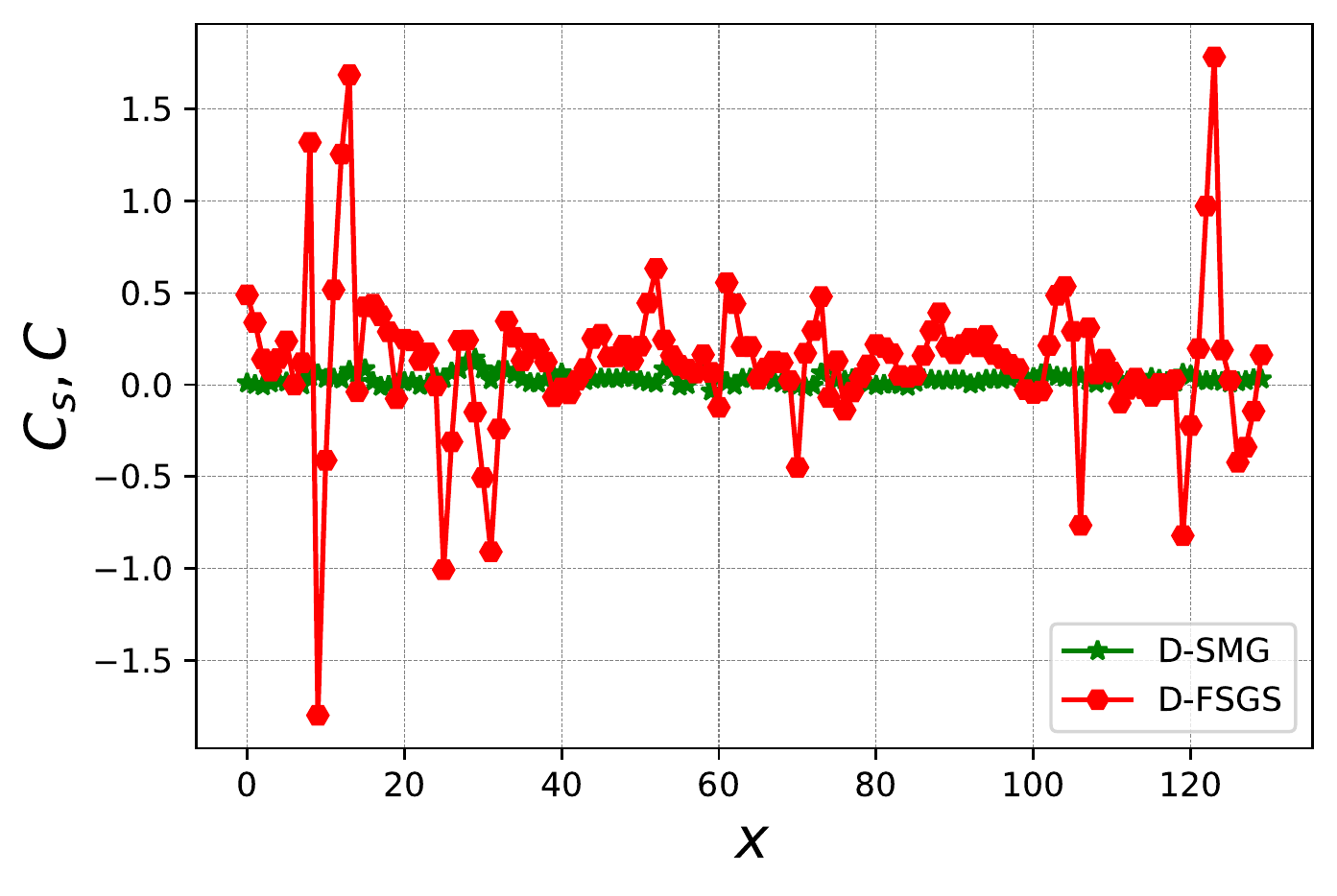}
    \caption{Comparing the model coefficients in dynamic Smagorinsky model (D-SMG) and dynamic FSGS  model (D-FSGS) in the middle imaginary line in a periodic domain using ${\mathcal{L}_{\delta}} = 4$.}\label{fig: Model_Constants}
\end{figure}

We have used Lilly’s approach for contracting purposes based on the Least Square Method that was described by Lilly in 1992  \cite{lilly1992proposed}. Therefore, one can write Eq. \ref{eq.tensorial} as
\begin{align}\label{eq.e}
       \big ((\nabla . G)_{i} - C N_i)^{2} = e, 
\end{align}
in which $e$ is the squared error and the nonlocal term $N_i$ is defined as
\begin{align}\label{eq.Ni}
N_i =  \big(  F(\alpha ) (-\Delta ^ {\alpha}) \widehat{\bar V_{i}} - \reallywidehat{ F(\alpha ) (-\Delta ^ {\alpha}) \bar V_{i}} \big).
\end{align}
Finding the unknown in minimized form is conveniently possible by putting the derivative equals to zero, considering and testing $\frac{\partial ^ 2 e}{\partial C^2} >0 $ for the error minimization scope. Finally, the scalar model coefficient $C$ can be computed dynamically as 
\begin{align}\label{eq:C }
           C = \frac{\langle (\nabla . G)_{i}  N_{i} \rangle} {\langle  N_{i} N_{i}  \rangle}. 
\end{align}
Numerical instability may be occurred due to the negative eddy-viscosity in prolonged periods of time. As a remedy, one can perform an averaging over the directions of statistical homogeneity as suggested by Germano et al. \cite{germano1991dynamic}. Figure \ref{fig: Model_Constants} illustrates the variations of model constants in D-SMG and D-FSGS models in the imaginary center-line of a periodic domain in the first direction to have a comparison between the model constant variations in the context of an example.
\section{\textit{A Priori} Analysis}\label{sec: APriori_Analysis}
We assess the performance of the D-FSGS and compare the results with the results of conventional LES models including SMG and D-SMG. Also, we discuss about the performance of the D-FSGS model using different filter sizes $\mathcal{L}_{\delta}$ (covering both the LES and VLES regions). Before starting the \textit{a priori} tests, we need to study the effects of the fractional-order parameter ($\alpha$) and its proper selection for different scenarios.

We utilize ten independent snapshots of forced HIT data distributed over enough eddy turnover times on a triply periodic domain, $\Omega = [0, 2 \pi]^3$, for each targeted Taylor Reynolds number. The covered Taylor Reynolds numbers are $170$, $240$, and $355$, and the corresponding grid resolution for each case is  $320^3$,  $520^3$, $1024^3$ , respectively. The turbulence dataset comes from our open-access pseudo-spectral parallel code, which has been elaborated in \cite{akhavan2020parallel}. 

\subsubsection{The Inference of Optimum Laplacian-Order $(\alpha^{opt})$}\label{subsec: Opt}
The D-FSGS model depends on the fractional-order, $\alpha$. We provide three different data-driven approaches for the calculation of this parameter.
\subparagraph{First Approach:}\label{subsec: Approach1} we perform a precise comparison between the obtained correlation coefficients, considering the effects of all stress components in the SGS tensor and the ground-truth stresses that come from the DNS results. Subsequently, we find the $\alpha ^{opt}$, where the maximum of $\rho_{ij} = \langle  Avg (\rho [ \tau^{DNS} _{ij}, \tau^{Model} _{ij} ] ) \rangle  $ is achieved, in which, $ \langle \cdot \rangle  $ denotes the ensemble-averaged on different snapshots of data, and for each snapshot of data, we consider the mean of all corresponding components of the SGS tensor.

\begin{figure}[t!]
    \begin{minipage}[b]{1\linewidth}
        \centering
        \includegraphics[width=.50\textwidth]{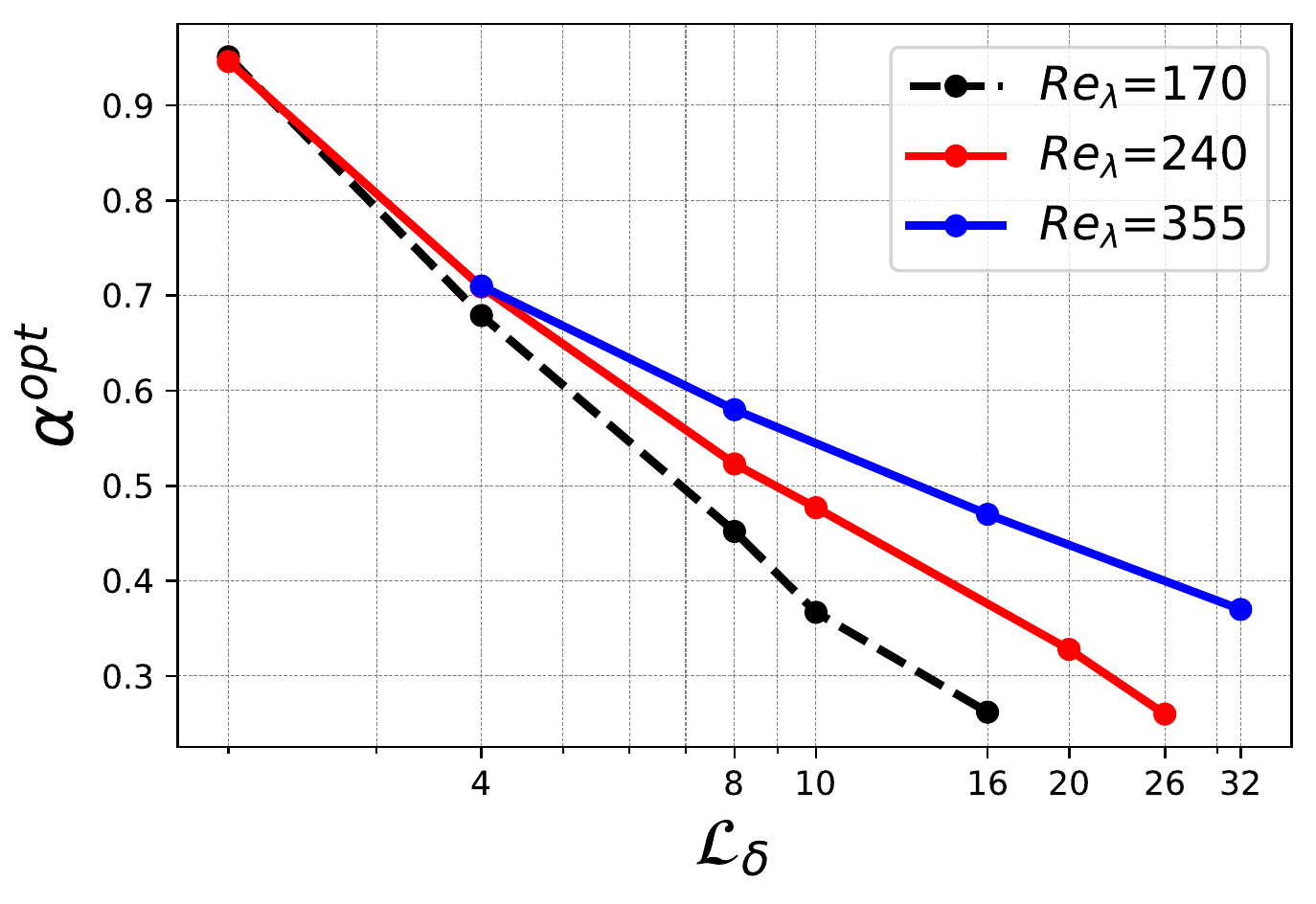}
        \subcaption{}\label{fig: opt_a}
    \end{minipage}
    \begin{minipage}[b]{.32\linewidth}
        \centering
        \includegraphics[width=\textwidth]{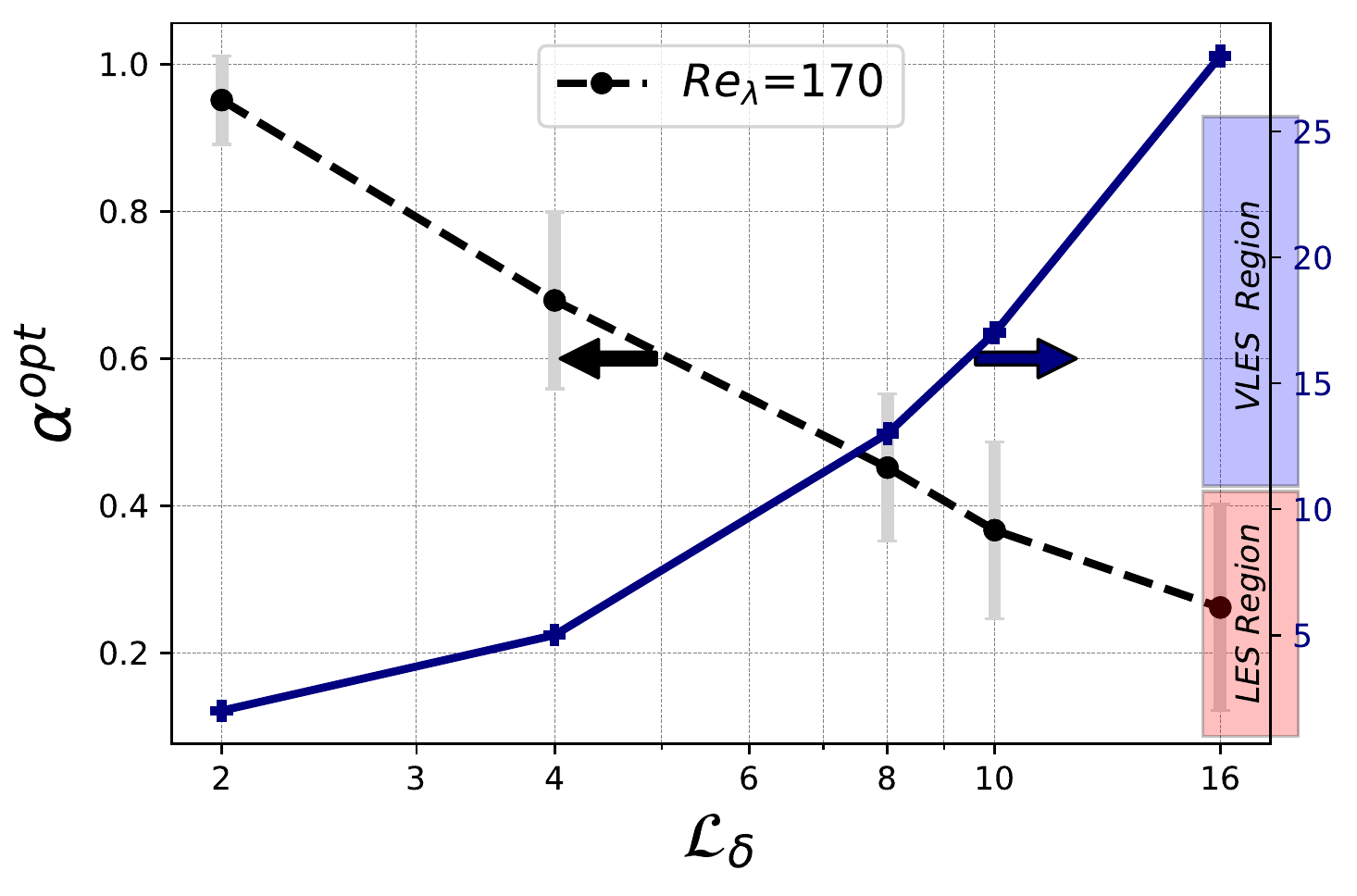}
        \subcaption{}\label{fig: opt_b}
    \end{minipage}
    \begin{minipage}[b]{.32\linewidth}
        \centering
        \includegraphics[width=\textwidth]{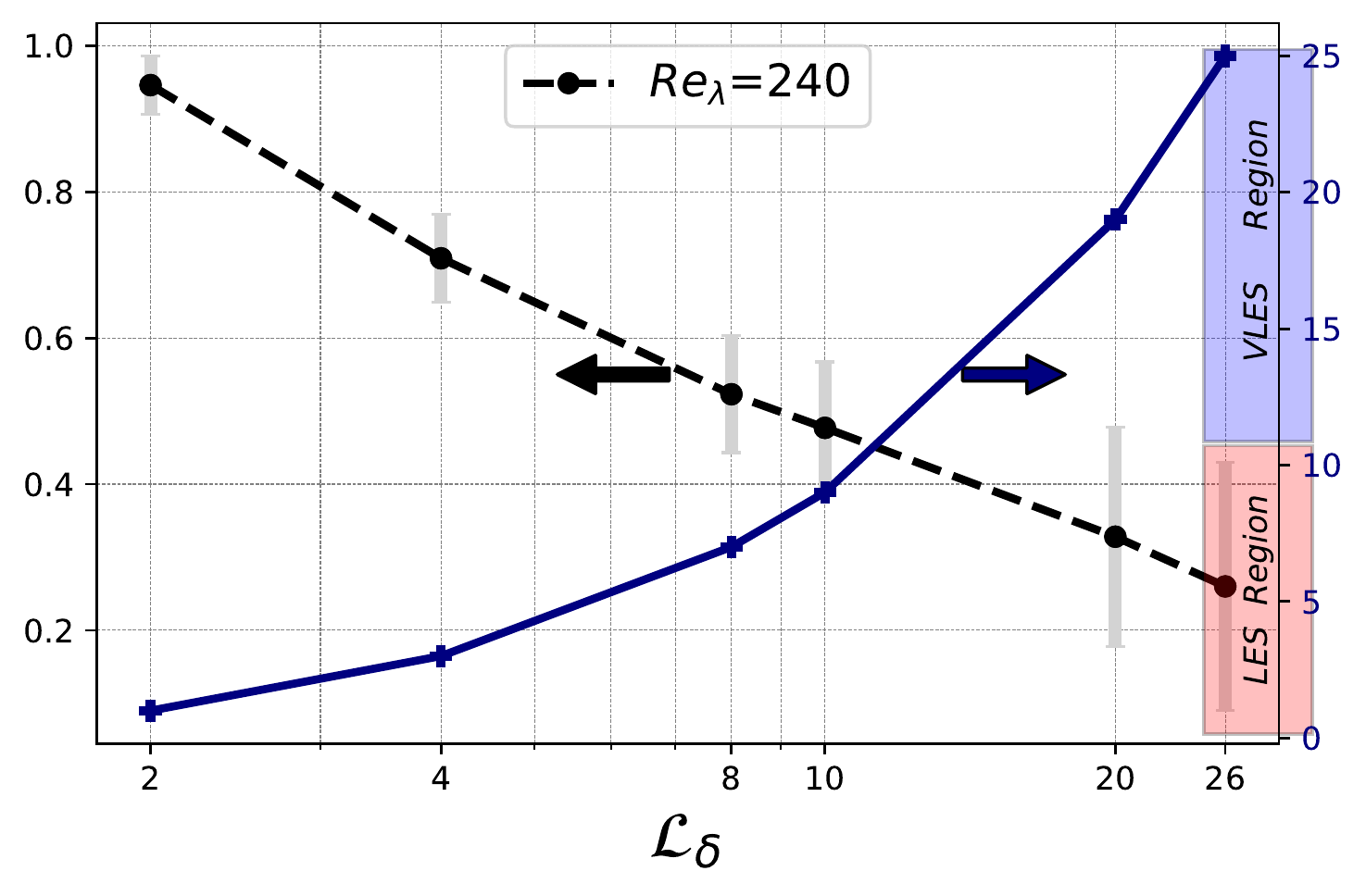}
        \subcaption{}\label{fig: opt_c}
    \end{minipage}
    \begin{minipage}[b]{.32\linewidth}
        \centering
        \includegraphics[width=\textwidth]{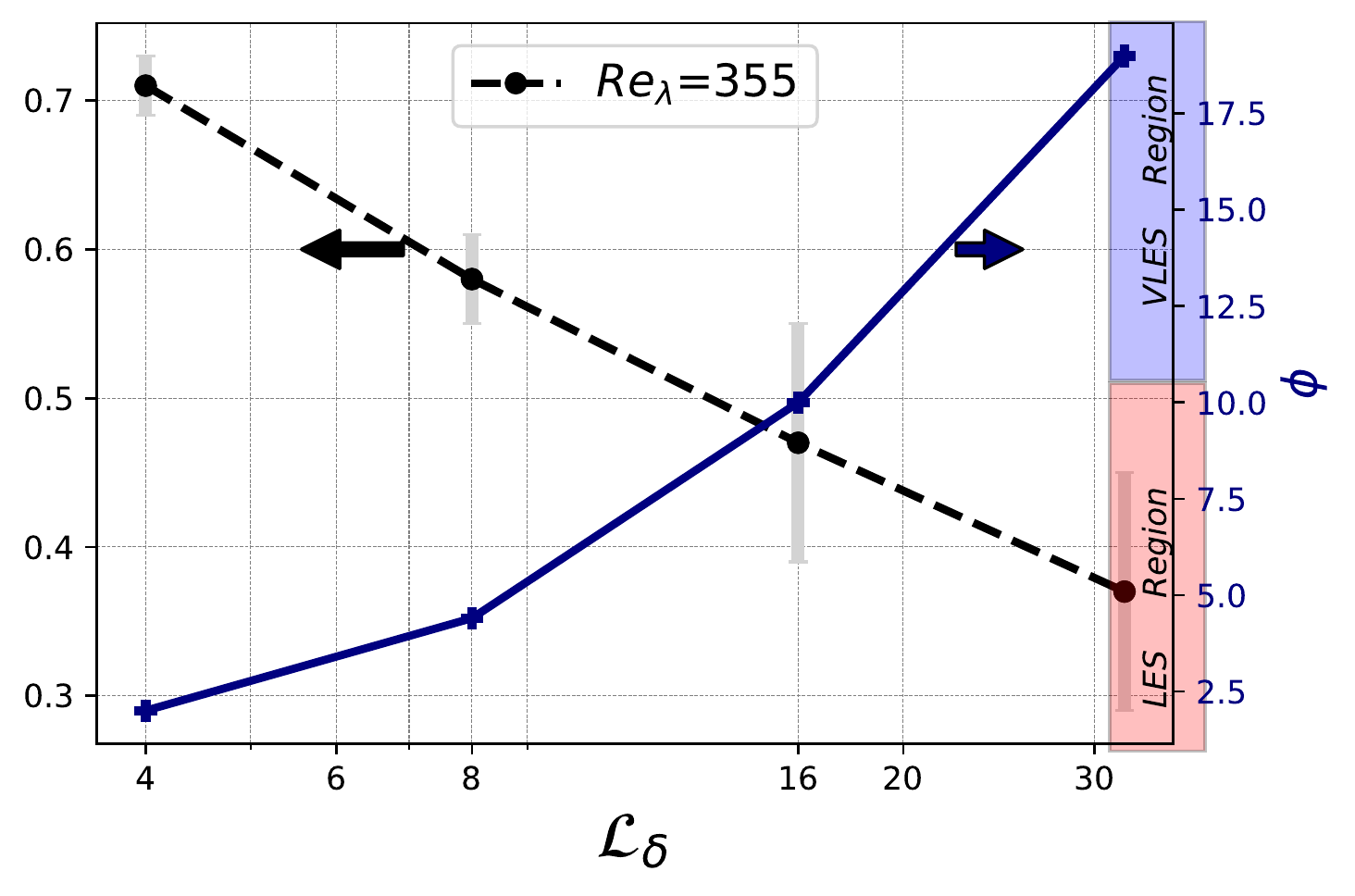}
        \subcaption{}\label{fig: opt_d}
    \end{minipage}
    \caption{Obtained optimum fractional-orders at different filter sizes based on the first approach, (a) effect of $Re_{\lambda}$,  amount of modeled turbulent kinetic energy when (b) $Re_{\lambda} = 170$, (c) $Re_{\lambda}=240$, and (d) $Re_{\lambda}=355$.}\label{fig: Optl,2,3}
\end{figure}

By sweeping the fractional-order values from zero to one and step size equals to 0.01, we determined the $\alpha ^{opt}$ for different characteristic filter sizes in three different Reynolds numbers. In Figure \ref{fig: Optl,2,3}, we have shown the $\alpha ^{opt}$ values for different filter sizes for three different Taylor Reynolds numbers. Also $\mathcal{L}_{\delta} = \frac{\mathcal{L}}{2 \delta_X }$, where $\delta_X = \frac{2 \pi}{N} $ shows the computational grid size associated with $N = 320, 520, 1024$ for three different Reynolds numbers. These results are obtained based on the ensemble-averaged quantities of ten snapshots of data for each Reynolds number to ensure meeting the necessary conditions regarding performing LES studies \cite{meneveau1994statistics}. There is an interestingly noticeable trend between the filter size and the obtained $\alpha ^{opt}$ from the ensemble-averaged relation for each Reynolds number (see Figure \ref{fig: opt_a}).
\subparagraph{Second Approach:}\label{subsec: Approach2} we employ a logarithmic regression to come up with a correlation between the $\alpha ^{opt}$  and ${\mathcal{L}_{\delta}}$ for each Reynolds number. The results are showing that $\alpha ^{opt}$ values obey a \textit{power-law} form, which reads 
\begin{align}\
    \label{eq.power-law}
           \alpha ^{opt} = a \ {\mathcal{L}_{\delta}} ^{b}.
\end{align}
\begin{table}[h]
	\caption{\footnotesize Proper parameters for being used in second approach (Eq. \ref{eq.power-law}) for determination of optimum Laplacian-order.}\label{tab: power-law_param}
	\centering
	\begin{tabular}{ccccc}
			\toprule \toprule
			$Re_{\lambda}$     &  $\quad$ & $a$    &  $\quad$ & $b$ \\
		\midrule
		170   &  $\quad$  & 1.53  &  $\quad$  & -0.62 \\
		240 &  $\quad$ & 1.39  &  $\quad$  & -0.49\\
		355  &  $\quad$  & 1.08  &  $\quad$  & -0.30\\
		\bottomrule \bottomrule 
	\end{tabular}
\end{table}
Increasing the filter size would incorporate more nonlocality, and that is the reason for the increase in the error bars in the VLES section; however, all cases converge statistically to the realized power-law relations. Moreover, there is a direct relationship between the Taylor Reynolds number and the optimum fractional-order in each filter size. Using the above relation for finding the $\alpha ^{opt}$ would be the second method that can be applied conveniently and confidently since the coefficient of determination, $R^2$, for these equations are above $0.98$, and indicates that the relation correlates very well with the measured data. The $a$, $b$ values in Eq. \ref{eq.power-law} are being determined for different taylor Reynolds numbers. Table \ref{tab: power-law_param} shows the proper parameters for being used in second approach of optimum Laplacian-order determination. The values in this table obtained based on the ensemble-averaged results of ten data snapshots for each $Re_{\lambda}$. Moreover, interpolation methods can be conveniently utilized using this table to find the proper value of $\alpha ^{opt}$ for other Reynolds numbers.

\subparagraph{Third Approach:}\label{subsec: Approach3}

we propose a third approach for obtaining the proper fractional-order for a quite wide range of (fine-to-very) LES region yet still in the \textit{a priori} sense, shown in Figure \ref{fig: Optl,2,3}.  The right vertical axes (secondary axis) in Figures \ref{fig: opt_b}, \ref{fig: opt_c}, \ref{fig: opt_d} show the volumetric averaged values of the modeled kinetic energy in percent ($\phi$) at that specific filter size. To calculate this value, $\phi$, we find the ratio between the resolved ($E_r$) and filtered ($E_f$) kinetic energy as 
\begin{align}\label{eq:Ef }
           &E_f = \frac{1}{2} \ \overline {u_i} \  \overline{u_i}, \\ \nonumber
           &E_r = \frac{1}{2} \ (\overline {u_i u_i} - \overline {u_i} \ \overline{u_i}).
\end{align}
Noting that almost all of the LES studies are designed to model about less than ten percent of the total kinetic energy, one can conclude based on the obtained plots for all $Re_{\lambda}$ numbers, taking  $\alpha = 0.5 $ can be an average corresponding factor for the LES region. We will show that even with this rough estimation, the proposed model provides better correlations in all of the stress components than the other conventional models. Therefore, using the second or third approach, we will have a dynamic nonlocal model, which is totally free of any tuning parameter including fractional-order. 
\subsubsection{Statistical Performance Assessment}\label{subsec: analysis}
We use ten three-dimensional snapshots of velocity fields, $u_i(x) , i= 1,2,3$, corresponding to $Re_{\lambda} = 240$ and perform \textit{a priori} testing to show the model performance. The DNS datasets with $Re_{\lambda} = 240$ and $N = 520$  will be utilized for the rest of the paper.
 \begin{figure}[t!]
    \begin{minipage}[b]{.51\linewidth}
        \centering
        \includegraphics[width=\textwidth]{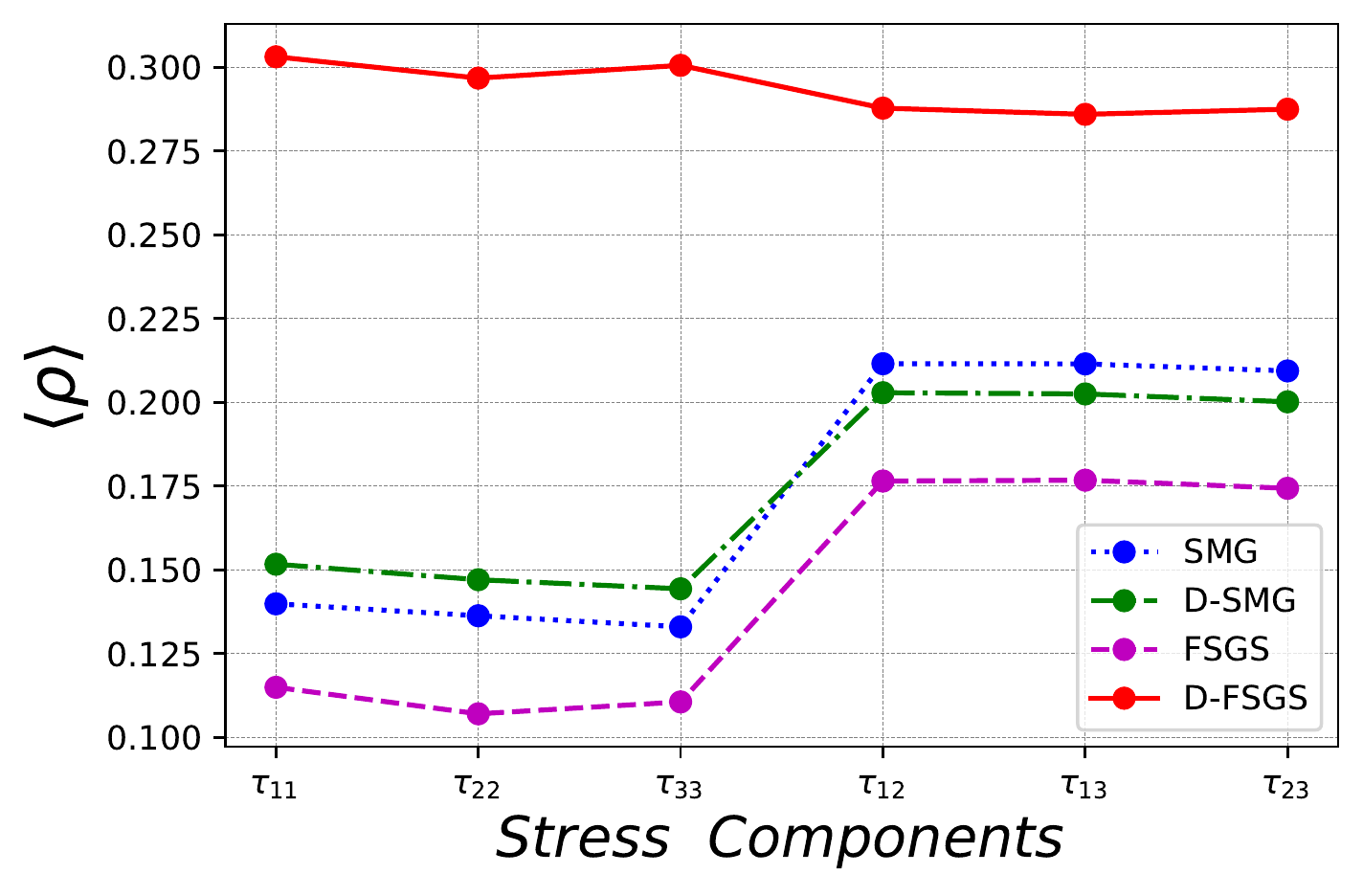}
        \subcaption{}\label{fig: Ens_corr1}
    \end{minipage}
    \begin{minipage}[b]{.47\linewidth}
        \centering
        \includegraphics[width=\textwidth]{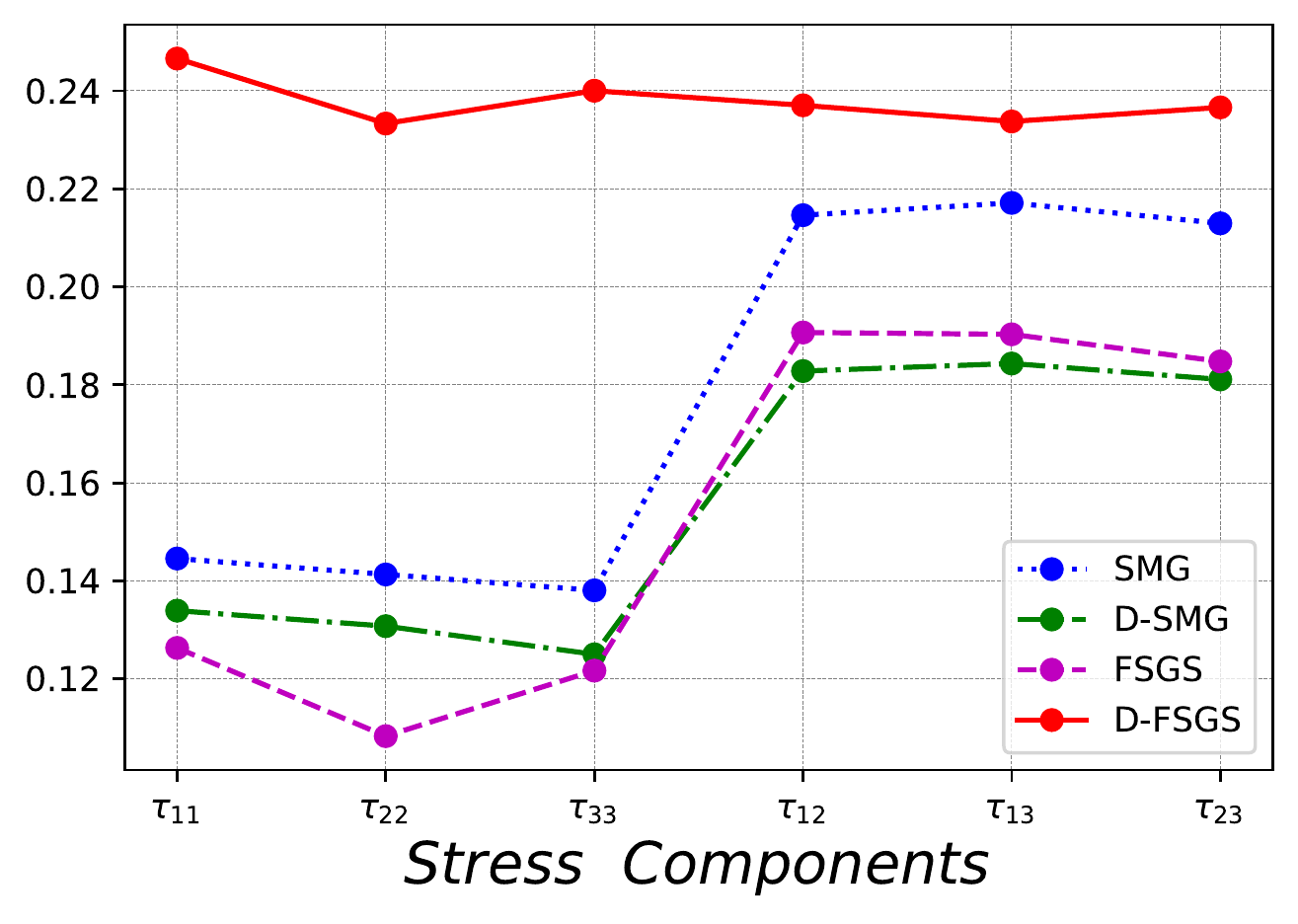}
        \subcaption{}\label{fig: Ens_corr2}
    \end{minipage}
    \caption{\textit{A priori} assessment in the context of the ensemble-averaged correlation coefficient comparisons in different models for (a) ${\mathcal{L}_{\delta}}$ = 4, and (b) ${\mathcal{L}_{\delta}}$ = 8 .}\label{fig: Ens_corr}
\end{figure}
We have depicted the ensemble-averaged correlation coefficients for all of the stress components in  Figure \ref{fig: Ens_corr} for the filter size equals to $4 $ and $8$. We see a remarkable better performance in both filter sizes and all directions for the new proposed model, D-FSGS. Interestingly, the coupling between the new dynamic procedure and the nonlocal nature of the model helps significantly to compensate for the directions in which we have low correlations. We see that D-FSGS outperforms static nonlocal (FSGS) and dynamic local (D-SMG) models in both filter sizes and all directions. Its well-known that, dynamic local models do not necessarily increase the correlation coefficient remarkably. In contrast to the D-SMG model, we see a significant raise in the correlation coefficient of the D-FSGS versus FSGS. Therefore, the new dynamic procedure not only adds the back-scatter prediction and being free from tuning constant capabilities, it also improve the performance as well. In another word, the dynamic procedure in the context of the nonlocal modeling seems to be more advantageous and fruitful than the local modeling.

\begin{figure}[t!]
    \begin{minipage}[b]{.505\linewidth}
        \centering
        \includegraphics[width=\textwidth]{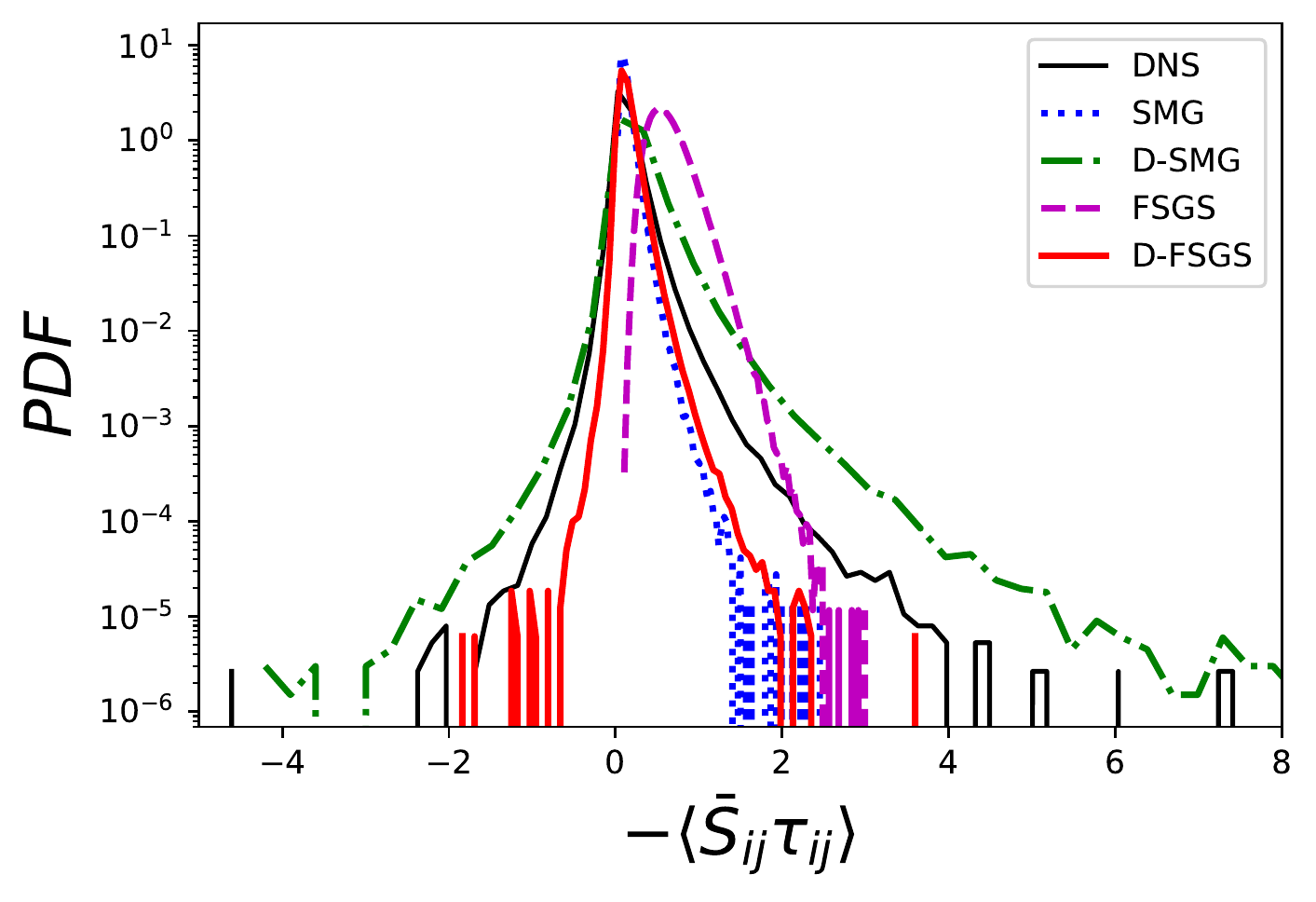}
        \subcaption{}\label{fig: Dissipation1}
    \end{minipage}
    \begin{minipage}[b]{.475\linewidth}
        \centering
        \includegraphics[width=\textwidth]{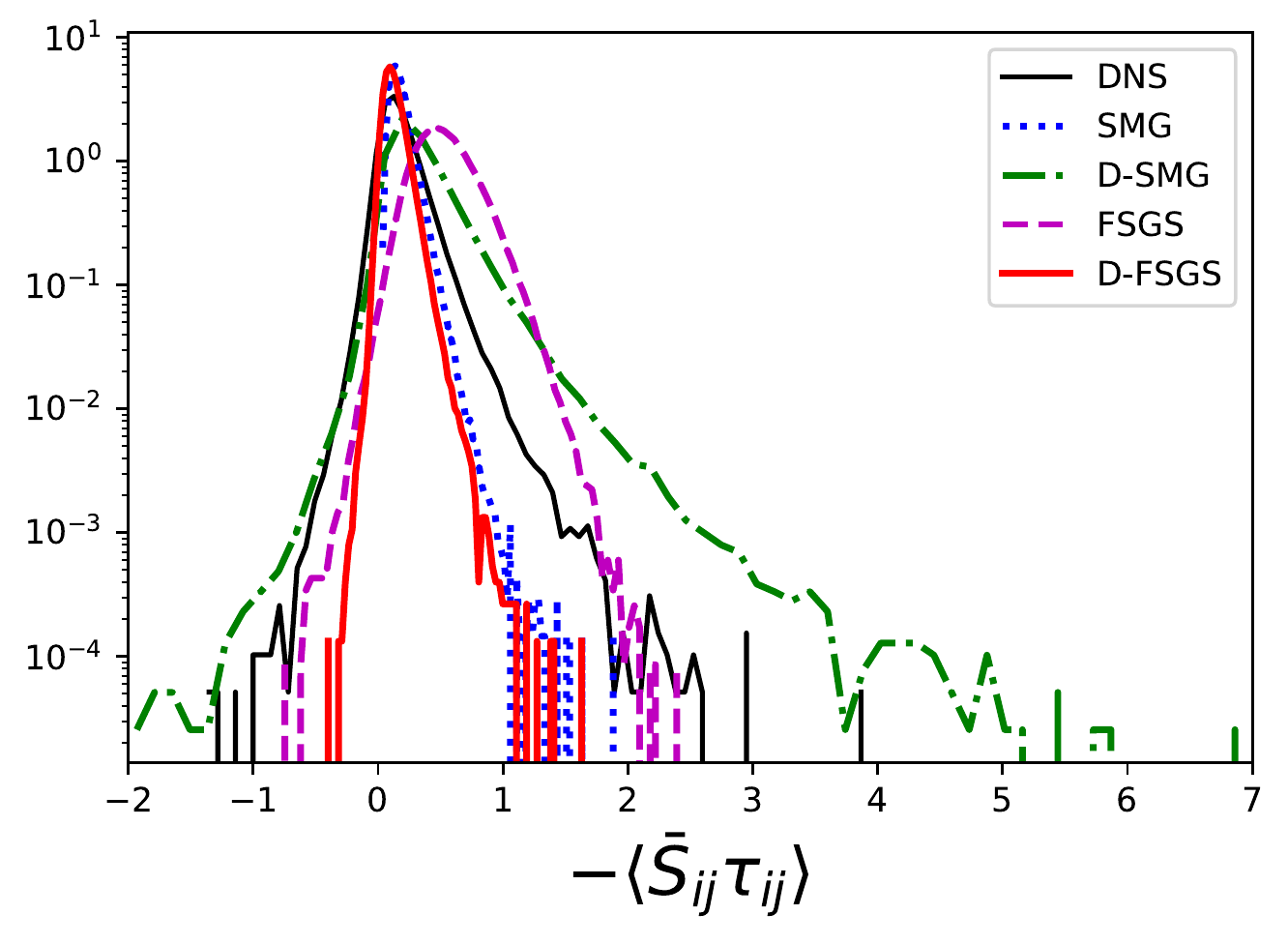}
        \subcaption{}\label{fig: Dissipation2}
    \end{minipage}
    \caption{PDF of ensemble-averaged SGS dissipation of kinetic energy using different models for (a) ${\mathcal{L}_{\delta}}$ = 4, and (b) ${\mathcal{L}_{\delta}}$ = 8.}\label{fig: Dissipation}
\end{figure}

In Figure \ref{fig: Dissipation} we have plotted the $\prod =-\langle \tau_{ij} \bar{S_{ij}} \rangle$, the \textit{SGS dissipation} of kinetic energy for the same filter sizes mentioned previously. The negative values of $\prod$ mean we have back-scatter, where the kinetic energy is transferred from subgrid to resolved scales. As the SGS dissipation of the filtered DNS results (ground-truth) show, a significant amount of flow shows back-scatter phenomena. We see the new model, D-FSGS, and D-SMG models are the only ones that truly predict the back-scattering in energy flow. Moreover, comparing the PDFs of dissipation show that the D-SMG model over-predicts the values for the tails of the distributions, which can be a potential factor in facing unstable conditions in simulations with this model necessitating the averaging operation in \textit{a posteriori} tests\cite{meneveau2000scale}. However, the new model provides closer results to the DNS ones. It should also be mentioned that a comparison of the percentage of grid points with back-scatter in filtered-DNS, D-SMG, and D-FSGS revealed that there is a better agreement for the D-FSGS model in the prediction of this quantity. As has been reported by Piomelli et al \cite{piomelli1991subgrid}, this percentage is a function of the filter type that one is using. In this study, three-dimensional box filtering for both the grid-level and test-level was utilized. The percentage of grid points with back-scattering in DNS was $26\%$, in D-SMG was $20\%$ and in D-FSGS was $24\%$ for ${\mathcal{L}_{\delta}} = 4$. For bigger filter size, ${\mathcal{L}_{\delta}} = 8$, these numbers were $30\%$, $18\%$ and $25\%$ for DNS, D-SMG and D-FSGS, respectively, which quantitatively indicates the better prediction of back-scatter for the D-FSGS model.
\begin{figure}[t!]
    \begin{minipage}[b]{.327\linewidth}
        \centering
        \includegraphics[width=1\textwidth]{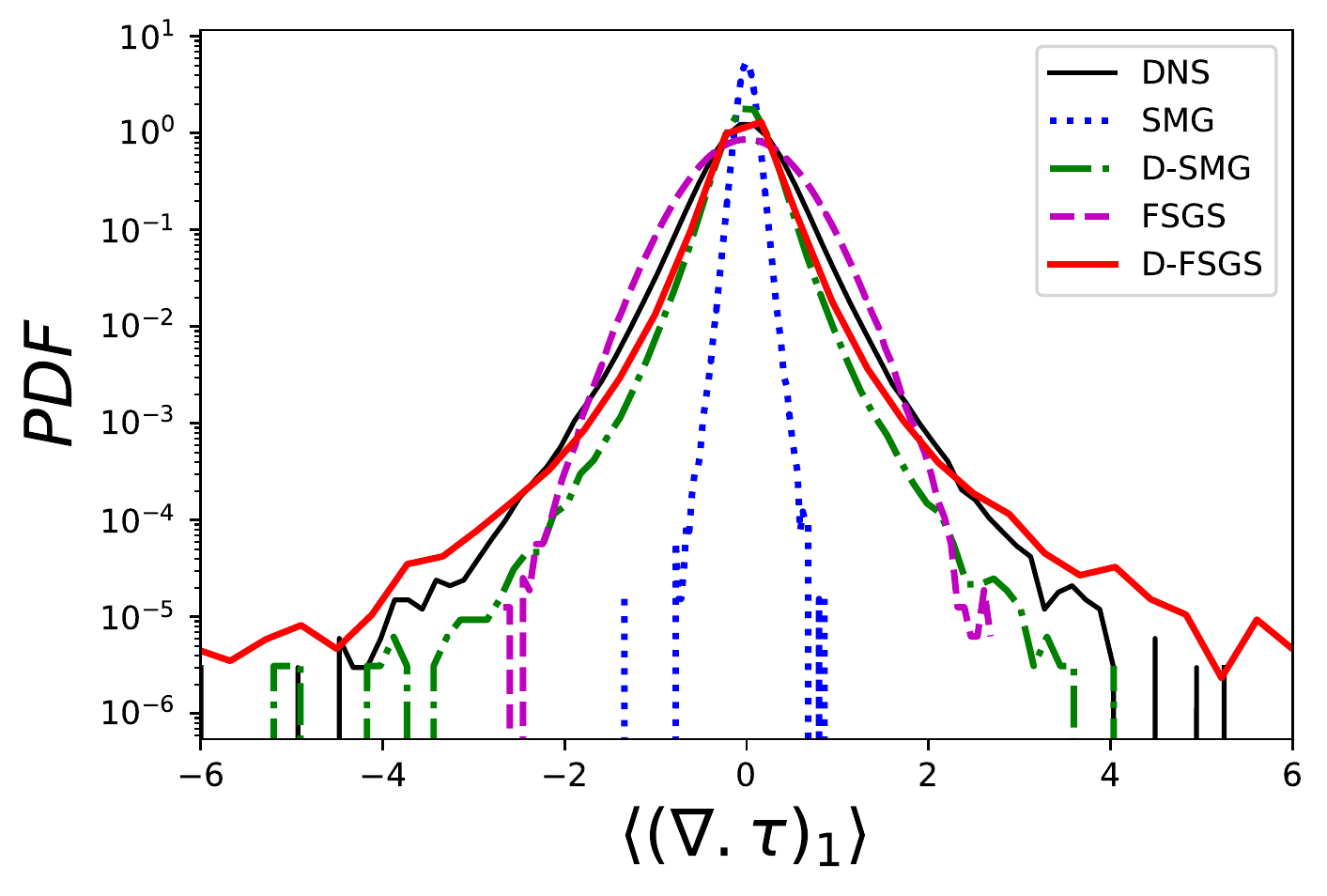}
        \subcaption{}
    \end{minipage}
    \begin{minipage}[b]{.31\linewidth}
        \centering
        \includegraphics[width=1\textwidth]{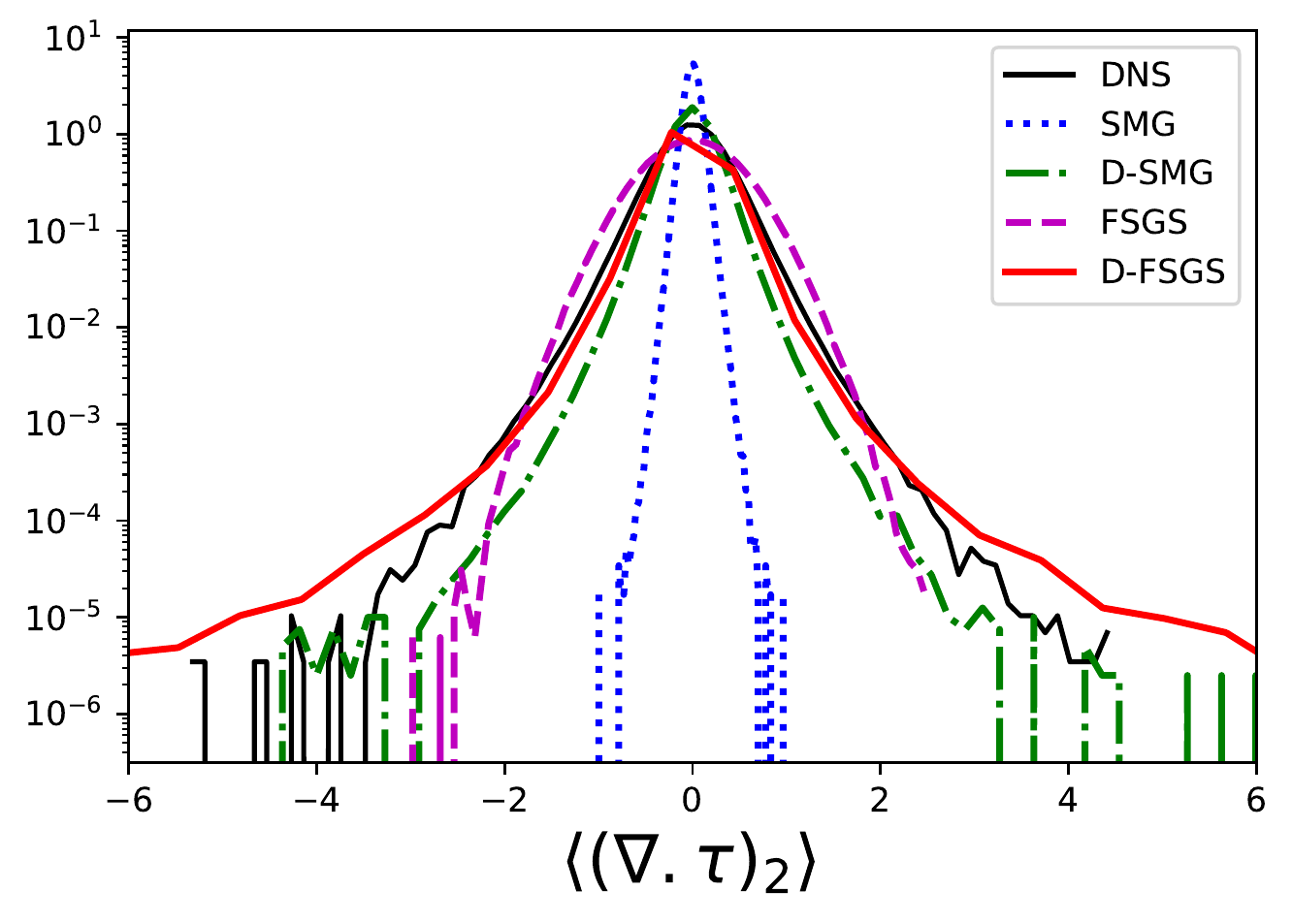}
        \subcaption{}
    \end{minipage}
    \begin{minipage}[b]{.31\linewidth}
        \centering
        \includegraphics[width=1\textwidth]{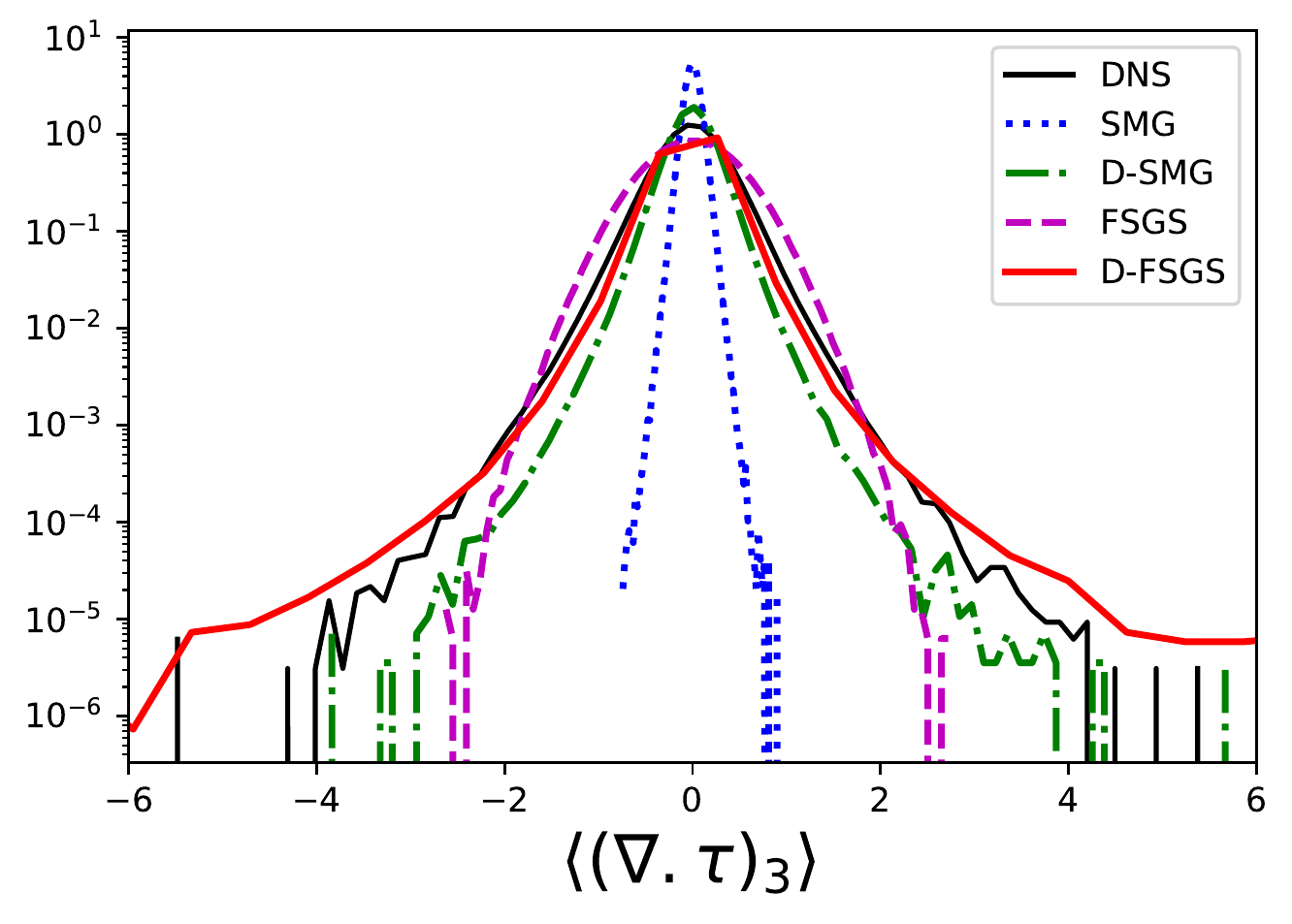}
        \subcaption{}
    \end{minipage}    
    \begin{minipage}[b]{.327\linewidth}
        \centering
        \includegraphics[width=1\textwidth]{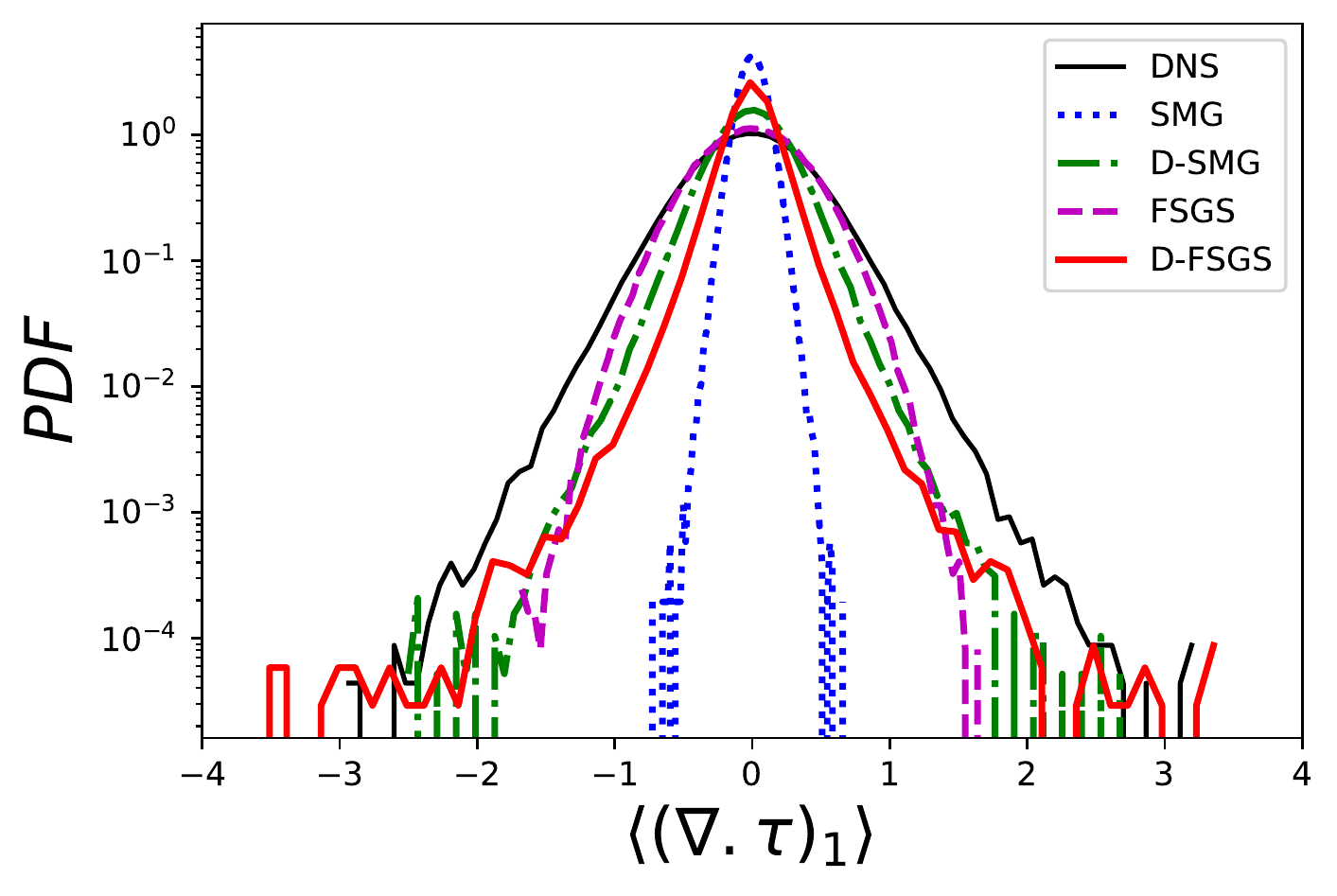}
        \subcaption{}
    \end{minipage}
    \begin{minipage}[b]{.31\linewidth}
        \centering
        \includegraphics[width=1\textwidth]{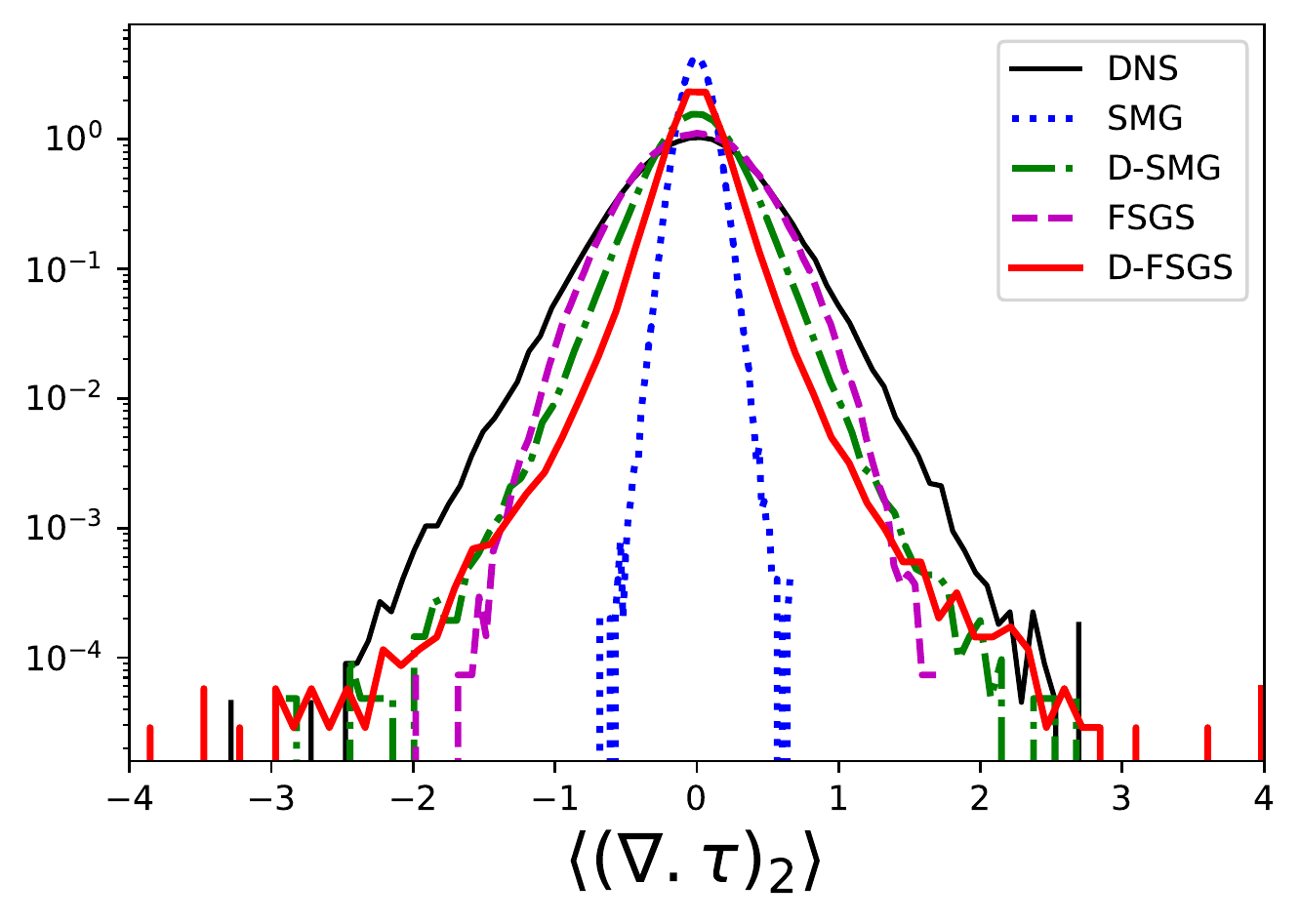}
        \subcaption{}
    \end{minipage}
    \begin{minipage}[b]{.31\linewidth}
        \centering
        \includegraphics[width=1\textwidth]{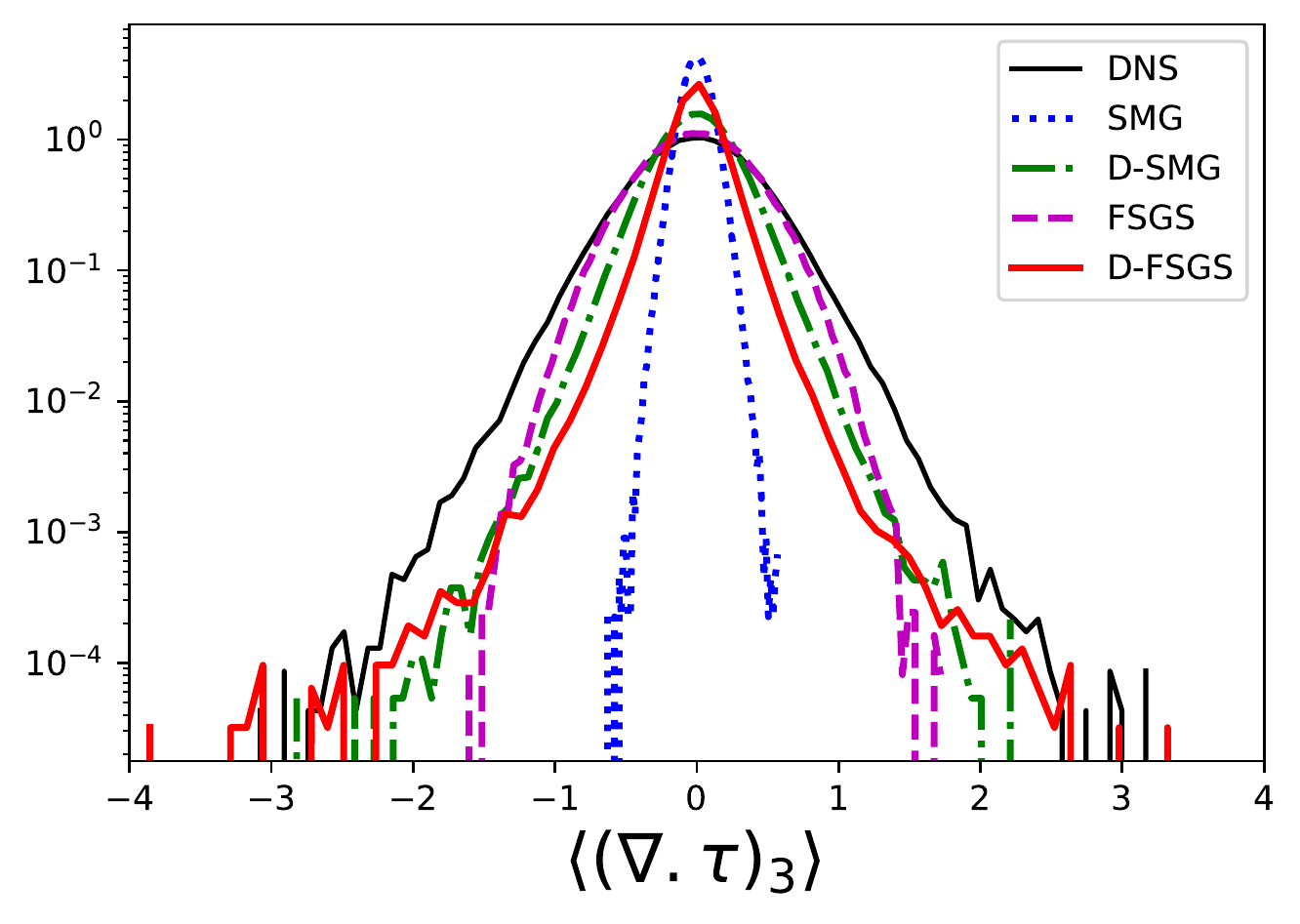}
        \subcaption{}
    \end{minipage}
    \caption{PDF of ensemble-averaged SGS forces using different models at  ${\mathcal{L}_{\delta}}$ =4 (first row), and  ${\mathcal{L}_{\delta}}$=8 (second row) .}\label{fig: SGS_forces}
\end{figure}
Figure \ref{fig: SGS_forces} shows the PDF of the ensemble-averaged SGS forces in three directions using ${\mathcal{L}_{\delta}} =4$ and ${\mathcal{L}_{\delta}} =8$. As it is clear in all three directions, there is superior math between the results of the D-FSGS model and DNS ones in comparison to the results of the Smagorinsky-based, and FSGS model considering tail and peaks of the SGS forces.
\begin{figure}[t!]
        \centering
        \includegraphics[width=0.6\textwidth]{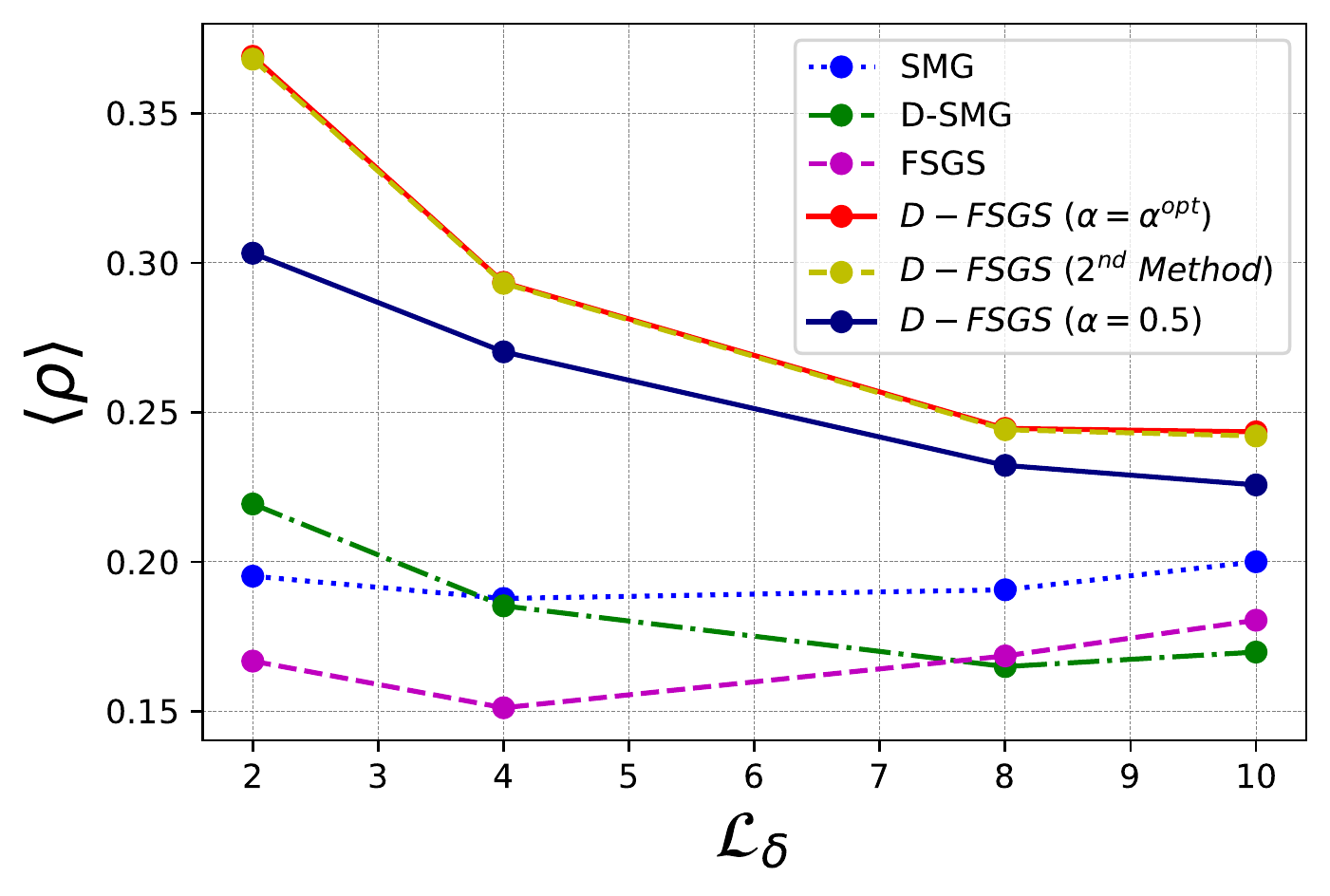}
    \caption{Model performance comparison using ensemble-averaged correlation at different filter sizes using three proposed approaches for the determination of the fractional-order.}\label{fig: Avg.corr}
\end{figure}
To employ the second and third approaches in the determination of the fractional-order, we do a test using $\alpha = \alpha^{opt}$ based on the precise determination method (first approach), then we apply the obtained power-law relations (second approach) to get $\alpha^{opt}$  and finally $\alpha = 0.5$ (third approach) to show its effect on the performance of the D-FSGS model. As has been depicted in Fig. \ref{fig: Avg.corr}, even using the rough-estimation approaches for $\alpha$, the D-FSGS model provides a higher correlation for all the filter sizes. Moreover, this plot again verifies the existence of power-law relation between the $\alpha^{opt}$ and filter size. One of the main disadvantages of nonlocal models in general and nonlocal turbulence models, in particular, is their sensitivity to the fractional-order. Surprisingly, the dynamic nonlocal model exhibits a less sensitivity to this parameter around the value $0.5$, since its effect is compensated by the automatic tuning parameter, thanks to the designed dynamic procedure. Moreover, the whole closure term in the LES framework indeed originates from filtering the first-order nonlinear convective term in the N-S equations. Here, the corresponding $\alpha^{opt} = 0.5$ in fact validates that the overall scaled up and nonlocal behavior of the closure term in our dynamic model emerges of total $2\alpha^{opt} = 1^{st}$ order.
\section{\textit{A Posteriori} Analysis}\label{sec: APosteriori_Analysis}
To assess the practical ability of the proposed dynamic nonlocal model to capture unsteady large-scale coherent structures and verifies numerical stability, we carry out the corresponding \textit{a posteriori} study here. In this section, we test the performance of the new proposed model (D-FSGS) against the static and dynamic Smagorinsky models (SMG, D-SMG) as well as the ground-truth filtered DNS results. For this purpose, pseudo-spectral N-S solver, which has been discussed in \cite{akhavan2020parallel}, was utilized for a triply periodic domain as $\Omega = [0, 2\pi]^3$. The resolution for this dataset is $520^3$  and large-eddy turnover times is $\tau_{\mathcal{L}} \simeq 2.7$. First, we set up a decaying HIT case in which the initial condition is based on the statistically stationary data sets. Thereafter, DNS and filtered-DNS results were gathered for the sake of comparisons. Second, each LES model including D-FSGS, SMG, D-SMG, and also unresolved numerical simulation (UNS), which is basically a DNS solver using the LES grid, implemented using the same pseudo-spectral solver, initiated with the filtered DNS initial condition for that specific filter size. In the mean time, the time step of the problems is chosen to have the CFL less than unity to ensure a stable time-integration. 
\begin{figure}[t!]
    \begin{minipage}[b]{.505\linewidth}
        \centering
        \includegraphics[width=\textwidth]{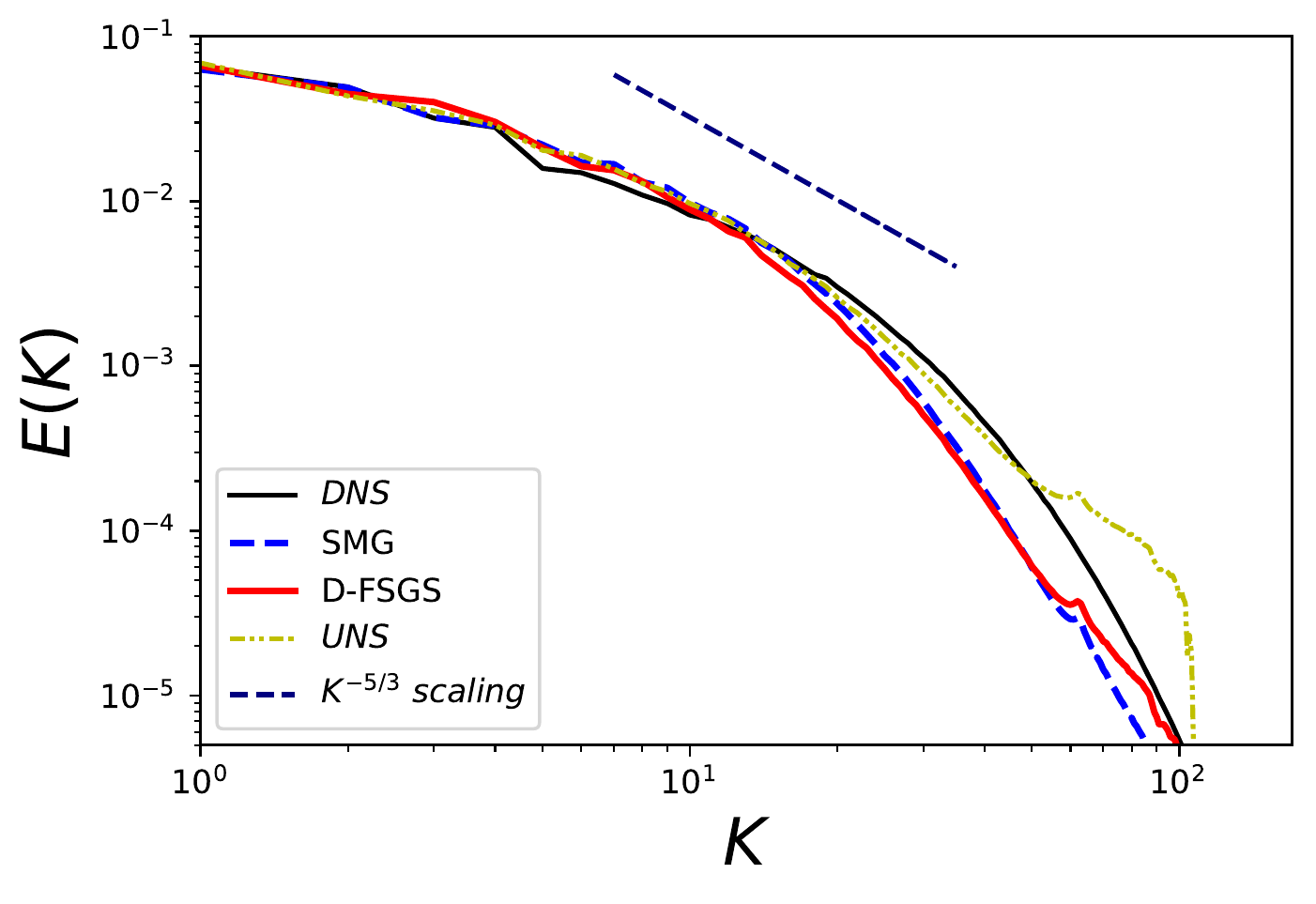}
        \subcaption{}\label{fig: Energy_Spectrum1}
    \end{minipage}
    \begin{minipage}[b]{.475\linewidth}
        \centering
        \includegraphics[width=\textwidth]{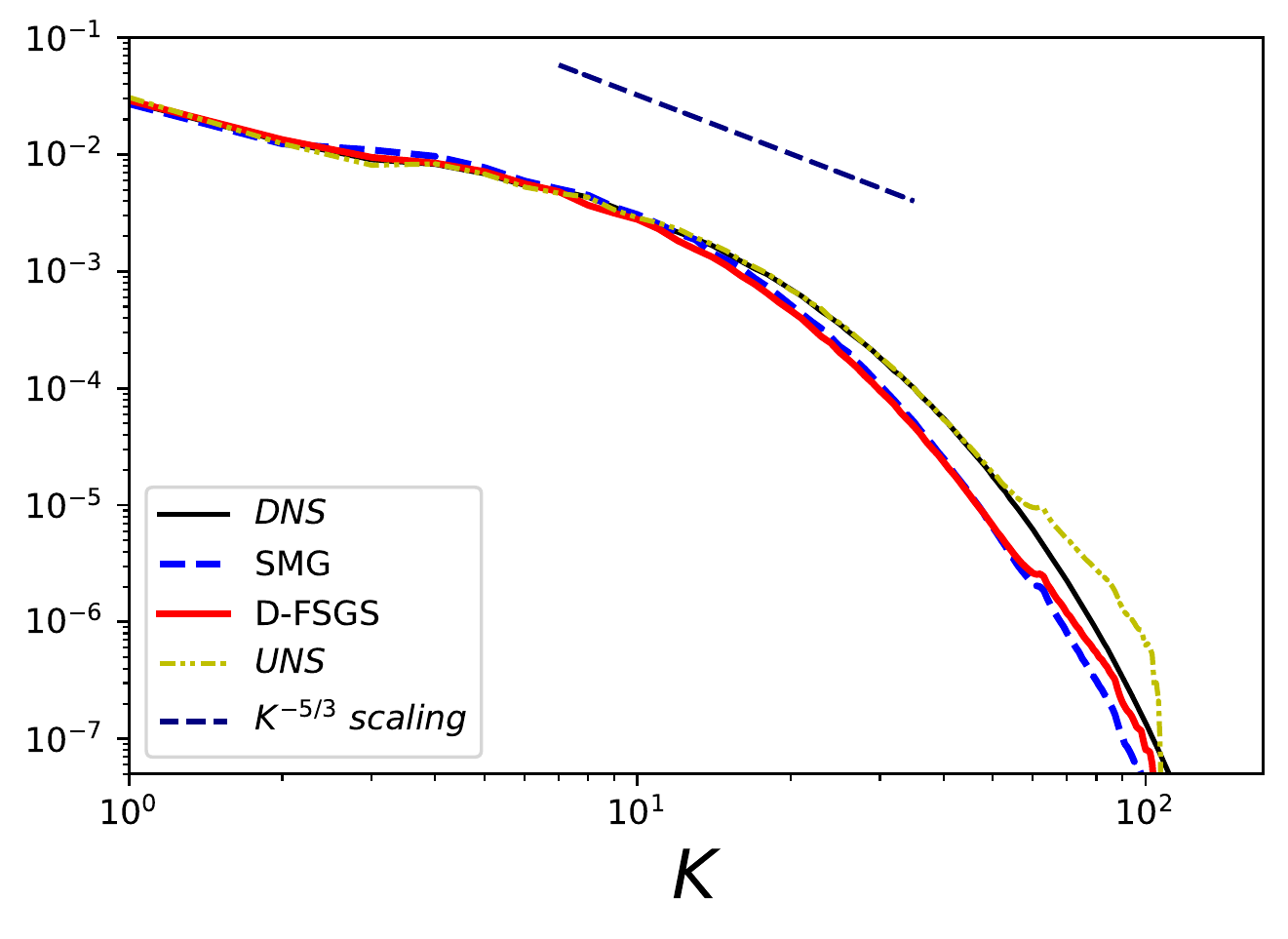}
        \subcaption{}\label{fig: Energy_Spectrum2}
    \end{minipage}
    \caption{Prediction of the kinetic energy spectra in different turbulence models at (a)  $\frac{t}{\tau_{\mathcal{L}}} \simeq 2$, and (b) $\frac{t}{\tau_{\mathcal{L}}} \simeq 4$ .}\label{fig: Energy_Spectrum}
\end{figure}
We compared the kinetic energy spectrum obtained from each model with the ground-truth ones. In this section we are just showing the results of using ${\mathcal{L}_{\delta}} = 4$ since in bigger filter widths, there is not enough resolution especially in the small scale sections. Fig. \ref{fig: Energy_Spectrum} illustrates energy spectrum in two relatively high integration times $\frac{t}{\tau_{\mathcal{L}}} \simeq 2, \ 4 $ after initiation of decay. Higher integration times were chosen to address the numerical stability of models, which can be a concern especially in the dynamic models. It is well noted that Smagorinsky model is one of the most dissipative models and this matter is evident in both plots. Moreover, the UNS case as one may expect should have an accumulation of noise due to the zero turbulent viscosity. Therefore, we can expect that the D-FSGS model should be ideally placed between highly dissipative SMG and UNS energy curves, which act as lower and higher bonds, respectively.

We continued the implementation of the \textit{a posteriori} analysis by comparing the performance of the models in decaying homogeneous isotropic turbulence with $Re_{\lambda} = 240$ using LES filter width ${\mathcal{L}_{\delta}} = 4, \ 20$ as representative of LES and VLES regions, respectively. Since the DNS and LES solvers have not been initiated with the same initial conditions, we are comparing the results of filtered-DNS and LES models.
\begin{figure}[t!]
    \begin{minipage}[b]{.505\linewidth}
        \centering
        \includegraphics[width=\textwidth]{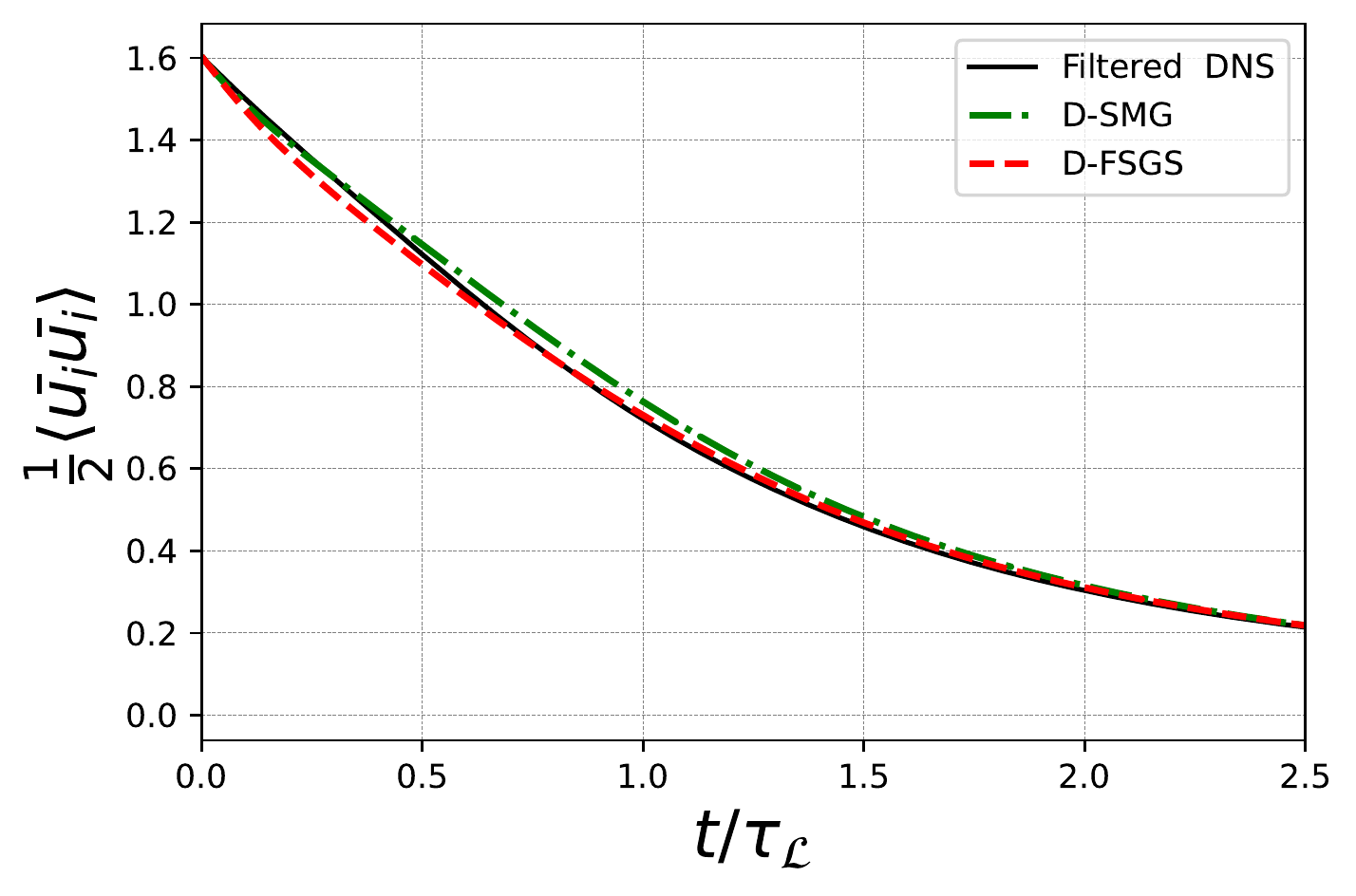}
        \subcaption{}\label{fig: Kinetic1}
    \end{minipage}
    \begin{minipage}[b]{.470\linewidth}
        \centering
        \includegraphics[width=\textwidth]{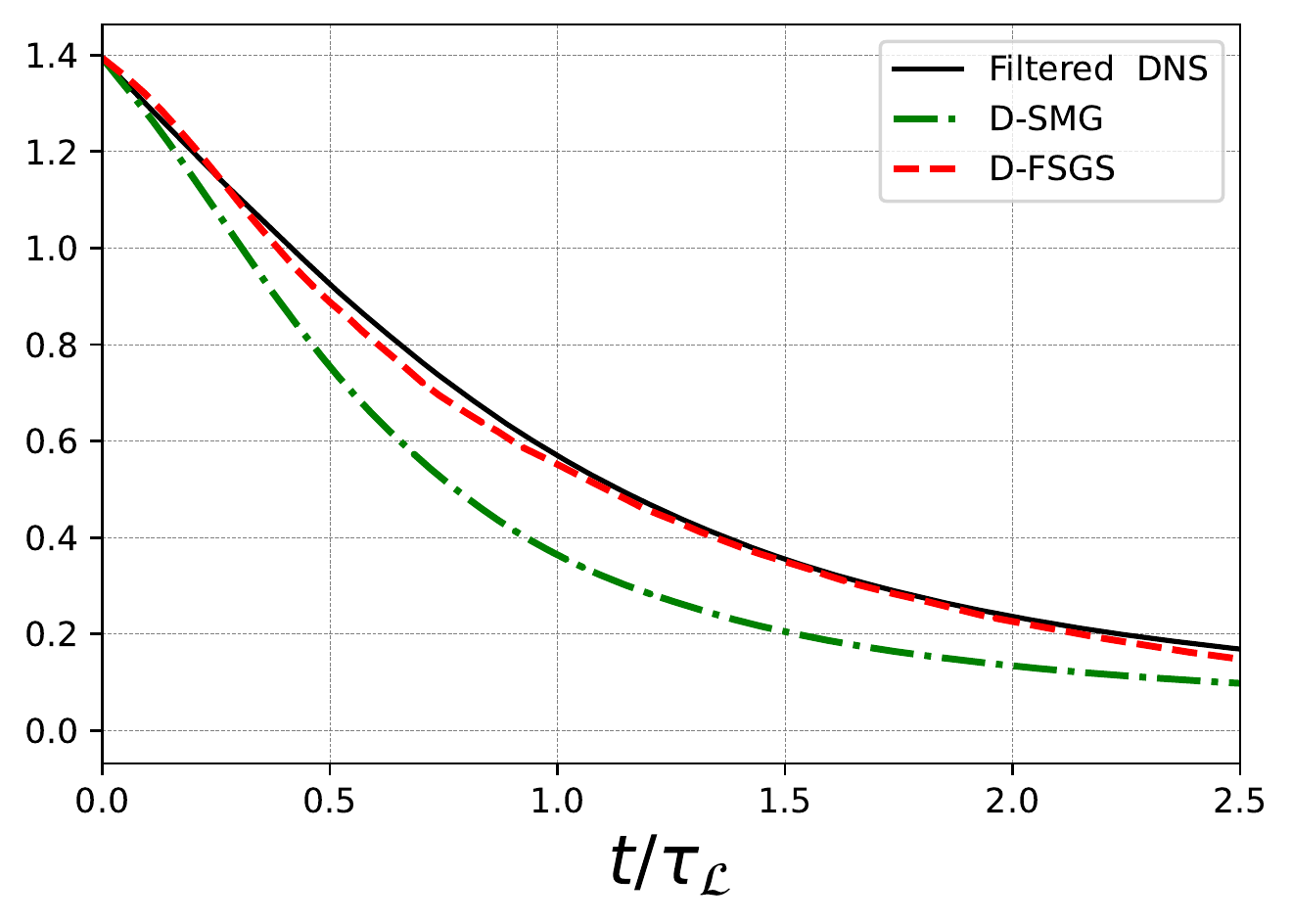}
        \subcaption{}\label{fig: Kinetic2}
    \end{minipage}
    \caption{Decay of the resolved turbulent kinetic energy in different turbulence models using (a) ${\mathcal{L}_{\delta}}=4$ as an LES case, and (b) ${\mathcal{L}_{\delta}}=20$ as a VLES case.}\label{fig: Kinetic}
\end{figure}
Fig. \ref{fig: Kinetic} shows the decay of resolved kinetic energy with time. As the plots show, the D-FSGS model provides better agreement with the filtered-DNS results in both LES and VLES cases. Interestingly, this difference is more remarkable in using bigger filter sizes and can be due to the fact that enlargement of the filter size incorporates more details to the solver and local models are not successful enough to correctly handling them due to the nature of the model. On the other hand, the D-FSGS model shows better performance by taking into account the nonlocal features and fractional derivations.
\begin{figure}[t!]
    \begin{minipage}[b]{.47\linewidth}
        \centering
        \includegraphics[width=\textwidth]{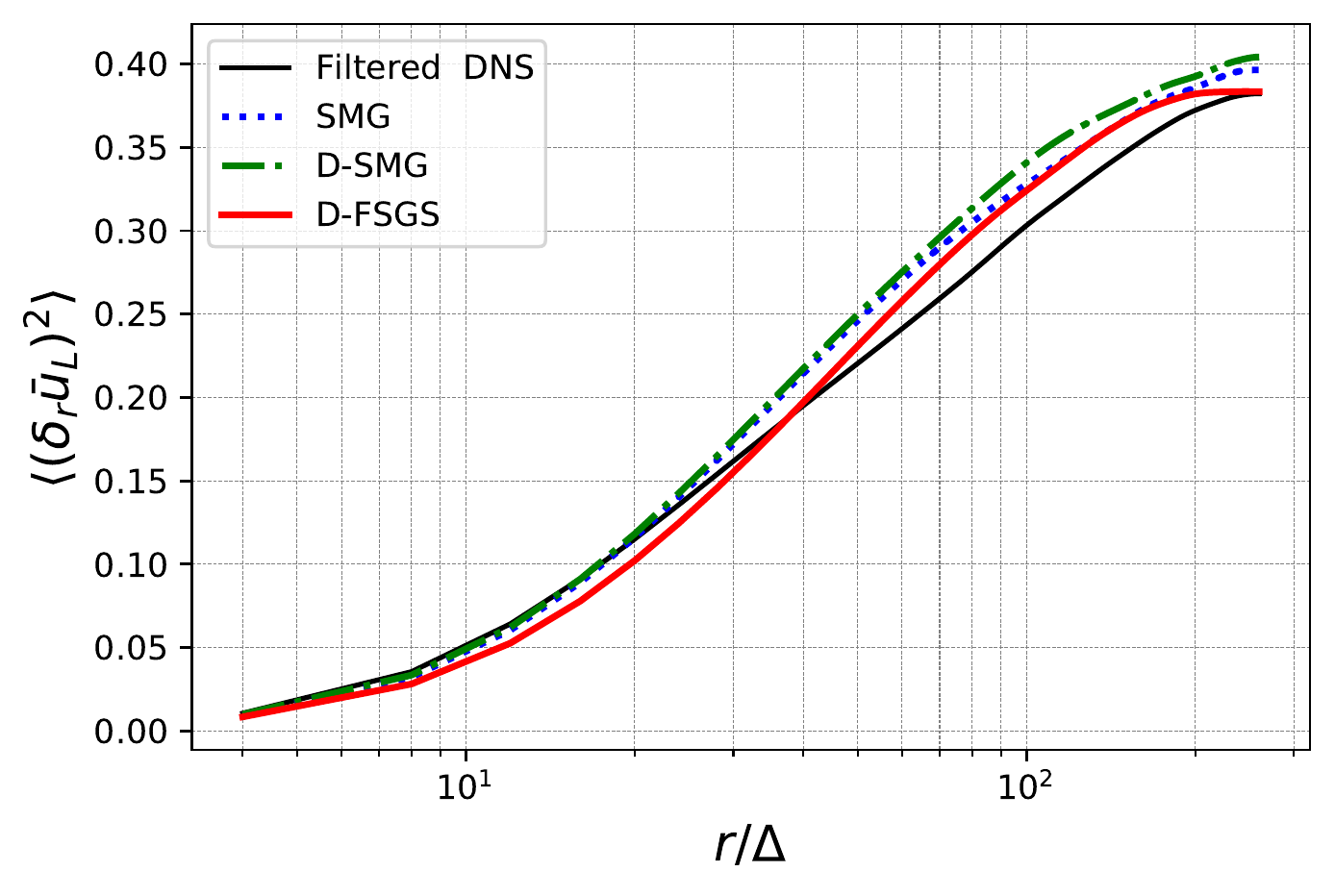}
        \subcaption{}\label{fig: Structure1}
    \end{minipage}
    \begin{minipage}[b]{.49\linewidth}
        \centering
        \includegraphics[width=\textwidth]{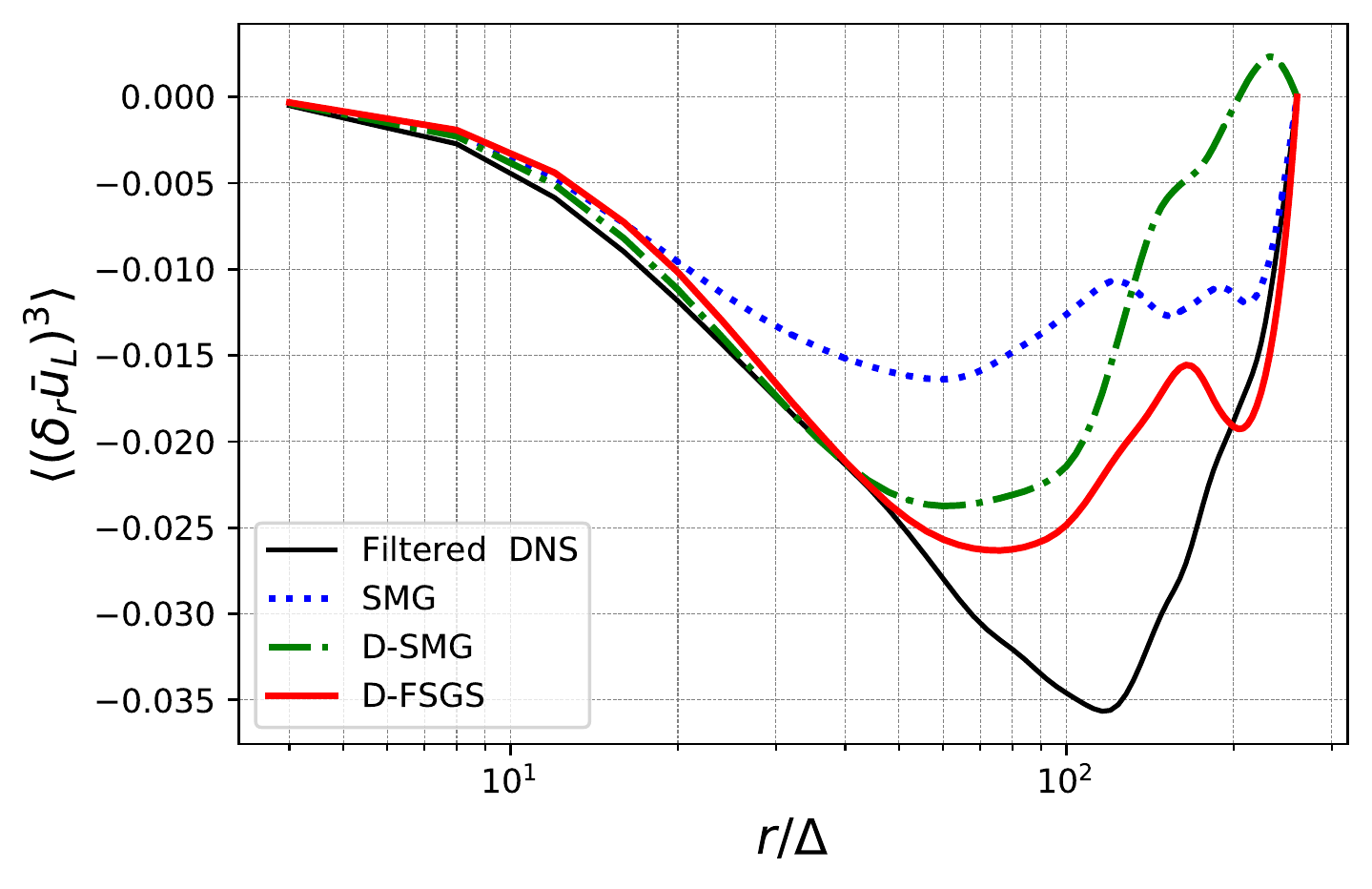}
        \subcaption{}\label{fig: Structure2}
    \end{minipage}
    \caption{Two-point diagnostics, velocity structure functions in different models and comparison to the filtered-DNS using ${\mathcal{L}_{\delta}} = 4$ at $\frac{t}{\tau_{\mathcal{L}}} \simeq 2$ for (a) n=2, and (b) n=3 in Eq. \ref{eq.structure}}\label{fig: Structure}
\end{figure}
As the last and most stringent test in the context of the two-point diagnostics, higher-order structure functions for the velocity increments are compared with the ground-truth filtered-DNS results. The second and third order structures functions are computed based on the following relation
\begin{align}\label{eq.structure}
    \langle (\delta _{r} \bar{u}_L )^n \rangle =  \langle \big[ \bar{u}_L (x+re_L) - \bar{u}_L (x) \big]^{n}  \rangle, \quad n = 2,3. 
\end{align}
To get the structure functions, first, we compute the velocity fields at a certain time for each model then we shift them based on ${u}_L (x+re_L)$, and finally, we do the filtering operation on each obtained field. The spatial shift is related to the filter size as $r = {\mathcal{L}_{\delta}} $. Comparing the results for the second and third-order structure functions in Fig. \ref{fig: Structure} for $\frac{t}{\tau_{\mathcal{L}}} \simeq 2$, reveals that all models are almost preserving the second-order structures, however, in the third-order structures we have more deviation for the Smagorinsky-based models. The new proposed model is more successful in the prediction of the third-order structures and its trend in the LES region.
\section{Concluding Remarks}\label{sec: Conclusion}
The present study introduces a novel dynamic nonlocal turbulence model for the isotropic turbulent flows, which can be applied for both LES and VLES purposes. The model has been developed based on the fractional Laplacian derivative of resolved velocity field, and a unique dynamic procedure defines the model constant in the divergence form. The effect of the fractional-order, $Re$ number, and characteristic filter size in LES and VLES cases are scrutinized using multiple high-fidelity and well-resolved DNS datasets belongs to forced homogeneous isotropic turbulence. Final decisions were made based on the ensemble-averaged quantities gathered from ten separate three-dimensional snapshots distributed over enough turnover times. The optimum fractional-order for each scenario is chosen when the maximum correlation is obtained for the ground-truth and predicted stresses of the model. The obtained results indicate that the relation between the filter size and fractional-order is closely obeying a power-law form for all the assessed $Re$ numbers. Also, there is a direct relationship between the $Re$ number and the fractional-order. Considering all the attained statistical results, two other data-driven straightforward methods for the determination of the fractional-order suggested and tested successfully. 

Our analysis included both \textit{a priori} and \textit{a posteriori} assessments. In the first one, we showed that there is a higher correlation between the results of the proposed method and the ground-truth DNS results in comparison to other conventional LES models. Also, the capability of the model in the prediction of the back-scatter is discussed in different filter sizes. In the \textit{a posteriori} assessments we tested the model performance in long-time integration in the context of a real N-S solvers, and its ability in capturing the large-scale coherent structures examined. Analysis were performed in a decaying isotropic turbulence scenario, and the new D-FSGS, static Smagorinsky and dynamic Smagorinsky models implemented separately. The results show that the new model is more successful in the prediction of the resolved turbulent kinetic energy in both LES and VLES studies. To test the models' performance in the long-time integration and its numerical stability, kinetic energy spectra are compared with the filtered-DNS results. Finally, two-point diagnostics were accomplished to compare different models' performance in the context of preserving second and third-order structures. The results demonstrate that the new model behaves better in preserving the high-order structures than the conventional ones. 

This study shows the promising potential of using bigger filter sizes rather than the conventional LES filter sizes. This important characteristic can be utilized to compensate the fractional model's high-cost demand. Moreover, accelerated evaluation methods for the fractional operators including learning-based approaches and fast solvers can be leveraged to pave the road for the real and more practical applications of nonlocal models.
\section*{Acknowledgement}
This work was supported by the ARO YIP award (W911NF-19-1-0444), and partially by the MURI/ARO award (W911NF-15-1-0562) and the NSF award (DMS-1923201). The HPC resources and services were provided by the Institute for Cyber-Enabled Research (ICER) at Michigan State University.
\section*{Data Availability}
The data that support the findings of this research are available from the corresponding author upon reasonable request.
\bibliography{ref}
\end{document}